\documentclass[prl,reprint,
superscriptaddress,showpacs,nofootinbib,%
tightenlines,twocolumn,floatfix
]{revtex4-1}
\usepackage{slashed}
\usepackage{epsfig}
\usepackage{color}
\usepackage{amssymb}

\newcommand{\ben}{\begin{displaymath}}
\newcommand{\een}{\end{displaymath}}
\newcommand{\be}{\begin{equation}}
\newcommand{\ee}{\end{equation}}
\newcommand{\bea}{\begin{eqnarray}}
\newcommand{\eea}{\end{eqnarray}}
\usepackage{bm}
\usepackage{amsmath,amssymb}
\newcommand{\tr}{\mathrm{tr}}
\newcommand{\I}{\mathrm{i}}
\def\Eq#1{Eq.~(\ref{#1})}
\def\Fig#1{Fig.~\ref{#1}}

\usepackage{array}

\begin{document}

\title{Strange electromagnetic form factors of the nucleon with $N_f = 2 + 1$ $\mathcal{O}(a)$-improved Wilson fermions}
\author{D.~Djukanovic}
 \affiliation{Helmholtz Institute Mainz, Staudingerweg 18, D-55128 Mainz, Germany}
\author{K.~Ottnad}
 \affiliation{Helmholtz Institute Mainz, Staudingerweg 18, D-55128 Mainz, Germany}
 \affiliation{PRISMA$^+$ Cluster of Excellence and Institute for Nuclear Physics, Johannes Gutenberg University of Mainz, Johann-Joachim-Becher-Weg 45, D-55128 Mainz, Germany}
\author{J.~Wilhelm}
 \affiliation{PRISMA$^+$ Cluster of Excellence and Institute for Nuclear Physics, Johannes Gutenberg University of Mainz, Johann-Joachim-Becher-Weg 45, D-55128 Mainz, Germany}
\author{H.~Wittig}
 \affiliation{Helmholtz Institute Mainz, Staudingerweg 18, D-55128 Mainz, Germany}
 \affiliation{PRISMA$^+$ Cluster of Excellence and Institute for Nuclear Physics, Johannes Gutenberg University of Mainz, Johann-Joachim-Becher-Weg 45, D-55128 Mainz, Germany}
\begin{abstract}
We present results for the strange contribution to the electromagnetic form
factors of the nucleon computed on the coordinated lattice simulation ensembles with $N_f=2+1$ flavors of
$\mathcal{O}(a)$-improved Wilson fermions and an $\mathcal{O}(a)$-improved
vector current. Several source-sink separations are investigated in order to
estimate the excited-state contamination. We calculate the form factors on six
ensembles with lattice spacings in the range of $a=0.049-0.086\,\text{fm}$ and
pion masses in the range of $m_\pi=200-360\,\text{MeV}$, which allows for a
controlled chiral and continuum
extrapolation. In the computation of the quark-disconnected contributions, we
employ hierarchical probing as a variance-reduction technique.
\end{abstract}


\maketitle

The contributions of strange sea quarks to the nucleon electromagnetic form
factors, which characterize the charge and current distribution in the nucleon,
have been of high interest in the last decades.  Experimentally, strange
electromagnetic form factors can be measured through the parity-violating
asymmetry, arising from the interference of the electromagnetic and neutral
weak interactions, in the elastic scattering of polarized electrons on
unpolarized protons. The first measurement by the SAMPLE experiment, at backward angles 
and low $Q^2$, yielded a result for $G_M^s$ which is
consistent with zero \cite{Spayde:2003nr}. The G0 collaboration combined
measurements at forward and backward angles and found a first indication of a
non-zero $G_E^s$ and $G_M^s$, contributing $\lesssim 10\%$ to the nucleon
electromagnetic form factors
\cite{PhysRevLett.104.012001,PhysRevLett.95.092001}. A first nonzero
measurement has been obtained by the A4 experiment at MAMI with a four momentum
transfer squared of $Q^2=0.22\,\text{GeV}^2$ , where $G_E^s = 0.050 \pm 0.038
\pm 0.019$ and $G_M^s = -0.14 \pm 0.11 \pm 0.11$ \cite{PhysRevLett.102.151803}.
A recent measurement from the HAPPEX collaboration at
$Q^2=0.624\,\text{GeV}^2$ found a value for the combination of
the strange electromagnetic form factors consistent with zero $G_E^s+0.517G_M^s
= 0.003\pm0.010\pm0.004\pm0.009$ \cite{PhysRevLett.108.102001}, confirming a previous
measurement at $Q^2=0.48\,\text{GeV}^2$, where a value consistent with
zero was found as well \cite{Aniol:2004hp}. For a recent review of the
experimental status of the strange electromagnetic form factors, see
\cite{Maas:2017snj}.
On the theoretical side, lattice QCD simulations allow for a
nonperturbative determination of the strange nucleon form factors. This is a
challenging calculation, due to the appearance of quark-disconnected diagrams,
which are notoriously difficult to evaluate. 
The most expensive part of the pertinent simulation is the calculation of the
trace of an all-to-all propagator. In order to obtain a good signal, the
application of  variance-reduction techniques, such as hierarchical probing
\cite{Stathopoulos:2013aci}, are crucial.
A prominent example to illustrate the importance of a precise knowledge of the
strange nucleon form factors is the weak charge of the proton. At tree level
and without radiative corrections, the weak charge is connected to the weak
mixing angle through $Q_W(p) = 1-4\sin^2\theta_W$. Hence, through measurements
of $Q_W(p)$, one can determine a fundamental parameter of the Standard Model. The
experiment proceeds by measuring the parity-violating asymmetry, from which
$Q_W(p)$ can be isolated, provided that the required nucleon form factors to describe the
hadronic contribution are known \cite{Becker:2018ggl,Maas:2017snj}. Here the strange
electromagnetic form factors $G_E^s$ and $G_M^s$, as well as the strange axial
form factor $G_A^s$, play a crucial role, as they constitute the leading uncertainty. 
In this Letter, we closely follow the strategy outlined in \cite{Djukanovic:2018iir}.

\begin{figure}[b]
	\includegraphics[scale=.7]{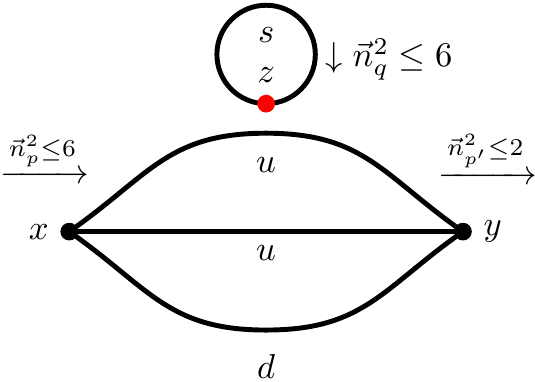}
\caption{Disconnected three-point function with a vector current inserted in
the strange loop (red dot). For the range of  momenta at the source and current
insertion, we use $\vec{n}_{p/q}^{2} \leq 6$, while at the sink, we restrict the
range to $\vec{n}_{p^{\prime}}^{2} \leq 2$ ($\vec{n}^2_{p/q/p^\prime}$ denote the units of squared lattice momenta). }
\label{fig:dis3pt}
\end{figure}

We make use of the coordinated lattice simulation (CLS) $N_f = 2+1$ $\mathcal{O}(a)$-improved Wilson fermion ensembles
with the tree-level-improved L\"uscher-Weisz gauge action \cite{Bruno2015}. The
fermion fields have open boundary conditions in time in order to prevent
topological freezing \cite{Luscher:2011kk}. Simulations have been performed
such that the sum of the bare quark masses is constant, which implies a
constant $\mathcal{O}(a)$-improved coupling \cite{BIETENHOLZ2010436}.
See Table~\ref{tab:ensembles} for a list of ensembles used in this Letter.
\renewcommand{\arraystretch}{1.5}
\begin{table*}[t]
\center
\begin{tabular}{cccccccccccc}
	&$\beta$	&$a$ [fm] &$N_s^3\times N_t$	&$m_\pi$\,[MeV]	&$m_K$\,[MeV]	& $m_N$\,[MeV]&$m_K L$ & $N_{\text{cfg}}$	&$N_{\text{meas}}$\\
\hline\hline
H105	&3.40	&0.08636 &$32^3\times 96$	&278	&460 &1037 & 6.44	&1020	&391680\\
N401$^*$	&3.46	&0.07634	&$48^3\times 128$	&289 & 462 & 1042 & 8.59	&701	&314048\\
N203	&3.55 &0.06426	&$48^3\times 128$ &345	&441 & 1111 & 6.90	&772&	345856\\
N200	&3.55 &0.06426	&$48^3\times 128$	&283	&463& 1061 & 7.23	&856	&383488\\
D200	&3.55 &0.06426	&$64^3\times 128$	&200	&480&989 &10.01	&278	&124544\\
N302$^*$	&3.70	&0.04981	&$48^3\times 128$	&354&	458 & 1120 &5.55	&1177	&527296\\
\end{tabular}
\caption{Gauge ensembles used in this Letter, where $N_{\text{cfg}}$ denotes the
number of gauge configurations and the last column corresponds to the total
number of measurements for the ratio in \Eq{eq:ratios}. The values for the
lattice spacing and pion and kaon masses are taken from \cite{Bruno:2016plf},
while the nucleon masses are estimated using the two-point function in this
work. For the ensembles marked with an asterisk, the pion and kaon masses have
been obtained from dedicated runs in connection with \cite{Ce:2018ziv}.}
\label{tab:ensembles}
\end{table*}
We obtain the strange electromagnetic form factors of the nucleon by
calculating the disconnected three-point function with a vector current
insertion in the strange quark loop.
The relevant diagram and our chosen momentum setup is depicted in
\Fig{fig:dis3pt}. The disconnected three-point function factorizes into
separate traces for the strange quark loop and the nucleon two-point function
\begin{multline}
C_{3,V_\mu}^{s}(\bm{q},z_0;\bm{p}^\prime,y_0,x;\Gamma_\nu) = \\
\Bigl\langle
e^{-\I\bm{q}\bm{x}} \mathcal{L}_{V_\mu}^{s}(\bm{q},z_0)\cdot
C_2(\bm{p}^\prime,y_0,x;\Gamma_\nu) \Bigr\rangle_G ,
\end{multline}
where $\mathcal{L}^s$ and $C_2$ denote the strange loop, given in Eq.
(\ref{eq:strange_loop}), and the nucleon two-point function respectively.

The calculation of nucleon two-point functions $C_2$ proceeds via
the standard nucleon interpolator
\begin{equation}
N_{\alpha}(x) = \epsilon_{abc}\left( u_\beta^a(x)\ \left(C\gamma_5\right)_{\beta\gamma}\ d_\gamma^b(x) \right)\ u_\alpha^c(x) ,
\end{equation}
and $\Gamma_0 = \frac{1}{2}(1+\gamma_0)$, which ensures the correct parity of the nucleon at zero momentum. Wuppertal
smearing \cite{Gusken:1989qx} is applied at the source and the sink for all
quark propagators. We increase the statistics of the nucleon two-point function
using the truncated solver method
\cite{Bali:2015qya,PhysRevD.91.114511}. 
Traces over the strange quark loops can be stochastically estimated using four-dimensional noise vectors $\eta$. For a local current 
\begin{align}
V^s&=\bar{s}(x)\Gamma s(x),
\end{align}
the trace over the strange quark loop then reads
\begin{figure*}[t]
\includegraphics[scale=.8]{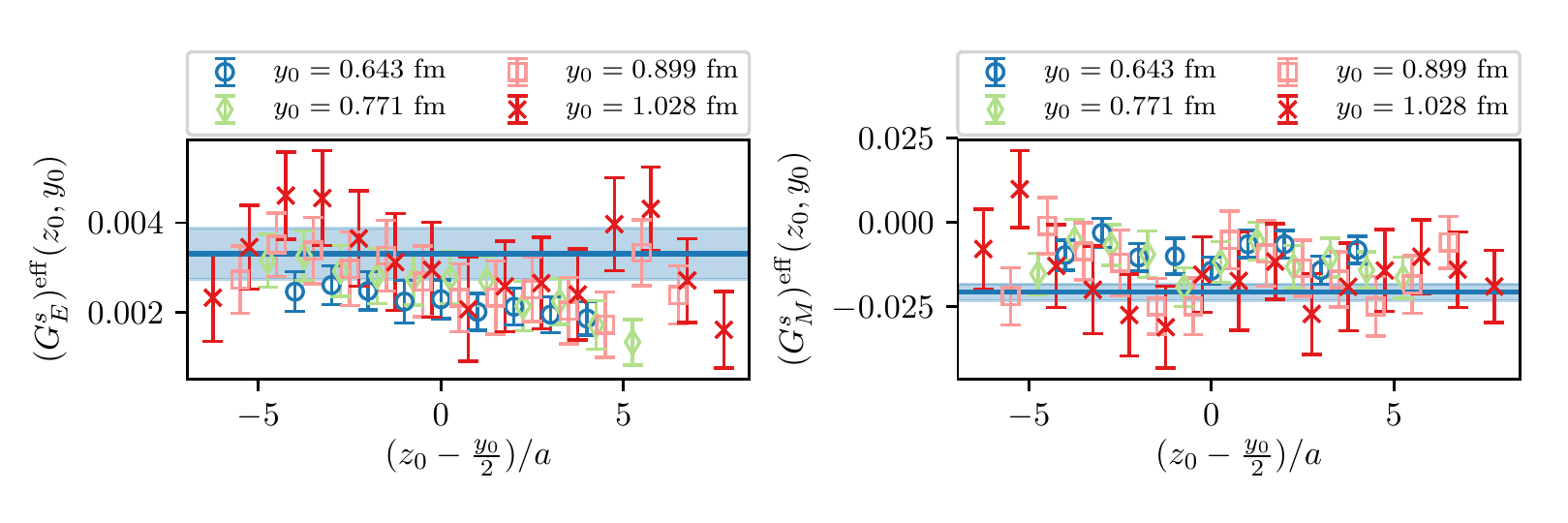}
\caption{
Results for the effective form factors on ensemble N200 determined via
Eq.~(\ref{eq:eff_sys}) at $Q^2=0.156\,\mathrm{GeV}^2$ compared to the estimate derived from the summation
method (horizontal band).}
\label{fig:N200esc}
\end{figure*}
\begin{equation}
\begin{split}
\left< \mathcal{L}_{\Gamma}^{s}(\bm{q},z_0) \right>_G &=
-\sum_{\bm{z}\in\Lambda} e^{i\bm{q}\cdot\bm{z}}\ \bigl<\tr\left[S^s(z;z)\
\Gamma\right]\bigr>_G \\&
= -\sum_{\bm{z}\in\Lambda} e^{i\bm{q}\cdot\bm{z}}\
\bigl<\eta^{\dagger}(z)\ \Gamma\ \psi(z)\bigr>_{G,\eta} ,
\label{eq:strange_loop}
\end{split}
\end{equation}
with 
\begin{align}
D^s \psi &= \eta,
\end{align} 
where $D^s$ denotes the Dirac operator for the strange quark, and the sum is
taken over the spatial volume $\Lambda$.  Instead of a local current we
consider the $\mathcal{O}(a)$-improved conserved vector current in this Letter
\begin{align}
\begin{split}
V_\mu(z)^{\text{Imp.}}=&\frac{1}{2}\Bigl( \bar{s}(z+\hat{\mu}a) (1+\gamma_\mu) U_\mu(z)^\dagger s(z)\\
& - \bar{s}(z) (1-\gamma_\mu) U_\mu(z) s(z+\hat{\mu}a)\Bigr)\\ 
& +ac_V\ \partial_\nu \left(\bar{s}(z) \sigma_{\mu\nu} s(z)\right) ,
\end{split}
\end{align}
with the improvement coefficient $c_V$ taken from \cite{PhysRevD.99.014519}. Furthermore, we use hierarchical
probing \cite{Stathopoulos:2013aci}, which replaces the sequence of noise
vectors by one noise vector multiplied with a sequence of Hadamard vectors. 
We find that the statistical error of the strange quark loop is reduced by a factor of 5 when using 512 Hadamard vectors, compared to the estimate based on 512 U(1) noise vectors, for nearly the same cost.
The quark
loops in this study were obtained by averaging two independent noise vectors
with 512 Hadamard vectors each. 
To extract the strange contribution to the electromagnetic form factors of the
nucleon, we consider the ratios (see \cite{Alexandrou:2008rp,Green:2014xba,Capitani:2015sba})
\begin{align}
R^s_{V_\mu}(z_0,\bm{q};y_0,\bm{p}^\prime;\Gamma_\nu) =
\frac{C^s_{3,V_\mu}(\bm{q},z_0;\bm{p}^\prime,y_0;\Gamma_\nu)}{C_2(\bm{p}^\prime,y_0)}\nonumber\\\times
\sqrt{\frac{C_2(\bm{p}^\prime,y_0)C_2(\bm{p}^\prime,z_0)C_2(\bm{p}^\prime\text{-}\bm{q},y_0\text{-}z_0)}{C_2(\bm{p}^\prime\text{-}\bm{q},y_0)C_2(\bm{p}^\prime\text{-}\bm{q},z_0)C_2(\bm{p}^\prime,y_0\text{-}z_0)}}
 .
\label{eq:ratios}
\end{align}
Performing the spectral decomposition and only taking the ground state into account,
these ratios read
\begin{widetext}
\begin{align} 
R^s_{V_\mu}(z_0,\bm{q};y_0,\bm{p}^\prime;\Gamma_\nu)
\xrightarrow[]{z_0,(y_0-z_0)\rightarrow \infty}
\frac{1}{4\sqrt{(E_{\bm{p}^\prime-\bm{q}}+m)(E_{\bm{p}^\prime}+m)E_{\bm{p}^\prime}E_{\bm{p}^\prime-\bm{q}}}}T\left(\widetilde{V}_{\mu}^s,\Gamma_\nu,\bm{q},\bm{p}^\prime\right)
,\\
T\left(\widetilde{V}_{\mu}^s,\Gamma_\nu,\bm{q},\bm{p}^\prime\right) = \ \tr\left[ \Gamma_\nu \left( E_{\bm{p}^\prime}\gamma_0 -i\bm{p}^\prime\bm{\gamma} + m \right) \widetilde{V}_{\mu}^s(\bm{q}) \left( E_{\bm{p}^\prime-\bm{q}}\gamma_0 -i(\bm{p}^\prime-\bm{q})\bm{\gamma} + m \right) \right] ,
\label{eq:trace}
\end{align}
\end{widetext}
where $\widetilde{V}_{\mu}^s$ can be obtained using the parametrization of the
nucleon matrix element 
\begin{equation}
\begin{split}
\left< N,\bm{k},s \left| V^\mu(x) \right| N,\bm{k}^\prime,s^\prime \right> = \bar{u}^{s}(\bm{k}) \Bigl( \gamma^\mu F_1(Q^2) \\+ \I\sigma^{\mu\nu} \frac{q_\nu}{2m}F_2(Q^2) \Bigr) u^{s^\prime}(\bm{k}^\prime)\ e^{\I q\cdot x} .
\end{split}
\end{equation}
We proceed by evaluating the trace in \Eq{eq:trace} for four different projectors
\be
\Gamma_0 = \frac{1}{2}(1+\gamma_0),\ \ \Gamma_{k} = \Gamma_0\ \I\gamma_5\gamma_k,\ \ k\in\{1,2,3\},
\ee
combined with all components of the vector current $\widetilde{V}_{\mu}^s$, leading to the asymptotic behavior of the ratios in the following form:
\begin{multline}
R^s_{V_\mu}(z_0,\bm{q};y_0,\bm{p}^\prime;\Gamma_\nu) \xrightarrow[]{z_0,(y_0-z_0)\rightarrow \infty}\\ M_{\mu\nu}^E(\bm{q},\bm{p}^\prime) G_E^s(Q^2) 
+ M_{\mu\nu}^M(\bm{q},\bm{p}^\prime) G_M^s(Q^2)\ .
\label{eq:Rkinfac}
\end{multline}
\renewcommand{\arraystretch}{1.0}
In analogy with Ref. \cite{Capitani:2017qpc}, we
collect all kinematic prefactors $M_{\mu\nu}^E$ and
$M_{\mu\nu}^M$ at a common $Q^2$ into a matrix $M$ and write the ratios as a
vector $\bm{R}$, which results in a (generally) overdetermined
system of equations for the form factors
$\bm{G}$
\begin{align}
\begin{split}
M &\bm{G} = \bm{R},\ \ 
M = \left( \begin{array}{c}
M^E_1\\
\vdots\\
M^E_N\\
\end{array}\ \begin{array}{c}
M^M_1\\
\vdots\\
M^M_N\\
\end{array} \right)  
,\ \ \\
&\bm{G} = 
\left(
\begin{array}{c}
G_E^s\\
G_M^s\\
\end{array}
\right),\ \ \bm{R} = 
\left(
\begin{array}{c}
R_1\\
\vdots\\
R_N\\
\end{array}
\right) . \label{eq:eff_sys}
\end{split}
\end{align}
The system can be solved by minimizing the least-squares function
\be
\chi^2 = (\bm{R}-M\bm{G})^T\ C^{-1}\ (\bm{R}-M\bm{G}) ,
\ee
where $C$ denotes the covariance matrix.
Note that we neglect all equations with vanishing kinematical factors ($M^E =
M^M = 0$) and average equivalent equations, i.e. with identical $M^E$ and
$M^M$. The latter average can already be carried out at the level of the nucleon
three-point functions, where the momenta of the nucleon states at the source
and the sink of the three-point functions are related by spatial symmetry
\cite{PhysRevD.81.034507}. In addition, averaging the nucleon two-point
functions over equivalent momentum classes, we construct the ratios in
\Eq{eq:ratios} from these averaged correlation functions. Solving the system of
equations at each $z_0$ and $y_0$ leads to the so-called effective form
factors, which still suffer from excited-state contamination.  Following
Refs.~\cite{Gusken:1989qx,Maiani:1987by,Doi:2009sq,Brandt:2011sj}, we obtain an
estimate of the asymptotic value of the form factors using the summation method
with source-sink separations in the range of $y_0=0.5-1.3\,\text{fm}$. 
In the case of the magnetic form factor, the plateau estimates show a clear trend towards the results obtained using the summation method. For the electric form factor, both methods agree already at small values of $y_0$.
The effective form factors for
several source-sink separations are shown in Fig.~\ref{fig:N200esc}. No significant deviation
from a plateau around the midpoint is visible. (We have included the
effective mass plot for the nucleon on  ensemble N200 in the Supplemental
Material \cite{supplmat:2019}.)

We will
use the summation method data as our standard dataset, since they are less
affected by excited-state contamination, compared to the plateau fits.
Nevertheless, we include the analysis of the plateau data, for a conservative
choice of source-sink separation of 1 fm using
5 points around the midpoint, as an estimate for the uncertainty coming from
excited states.  
In order to further analyze the kaon
mass and lattice spacing dependence, we use model-independent $z$-expansion fits
\cite{PhysRevD.82.113005,PhysRevD.90.074027} to fifth order to extract the
radii and magnetic moment. (We have explicitly checked that going to
a maximum order of 10 does not change the fit results.) The form factors can be expanded as
\begin{align}
\begin{split}
G_{E/M}(Q^2) = \sum_{k=1/0}^{5} a^{E/M}_k z(Q^2)^k, \\\ \ z(Q^2) = \frac{\sqrt{t_{\text{cut}}+Q^2}-\sqrt{t_{\text{cut}}}}{\sqrt{t_{\text{cut}}+Q^2}+\sqrt{t_{\text{cut}}}} .
\end{split}
\end{align}
Since the physical $\omega$ and $\phi$ mesons are narrow resonances and because
one cannot easily establish whether or not they are unstable particles on the
analyzed ensembles, we use $4 m_K^2$ for the value of the cut in the
$z$-expansion, where we use the ensemble kaon mass for $m_K$ (see Table~
\ref{tab:ensembles}).  We stabilize the fits using Gaussian priors centered
around zero for all coefficients with $k>1$. To this end, we first determine the
coefficients $a_{0,1}$ from a fit without priors and subsequently use the
maximum of these coefficients to estimate the width of the priors, i.e.,
$a_{k>1}=0\pm c \times \max\{|a_0|,|a_1|\}$.  We find that for $c=5$ the
extraction of the  radii and the magnetic moment are stable and lead to
consistent results even after applying a cut of $Q^2< 0.5$ GeV$^2$. 
Finally, we estimate the effect of this choice on
the final observables by repeating the analysis with the prior width doubled.
\begin{figure}[t]
\includegraphics[scale=.6]{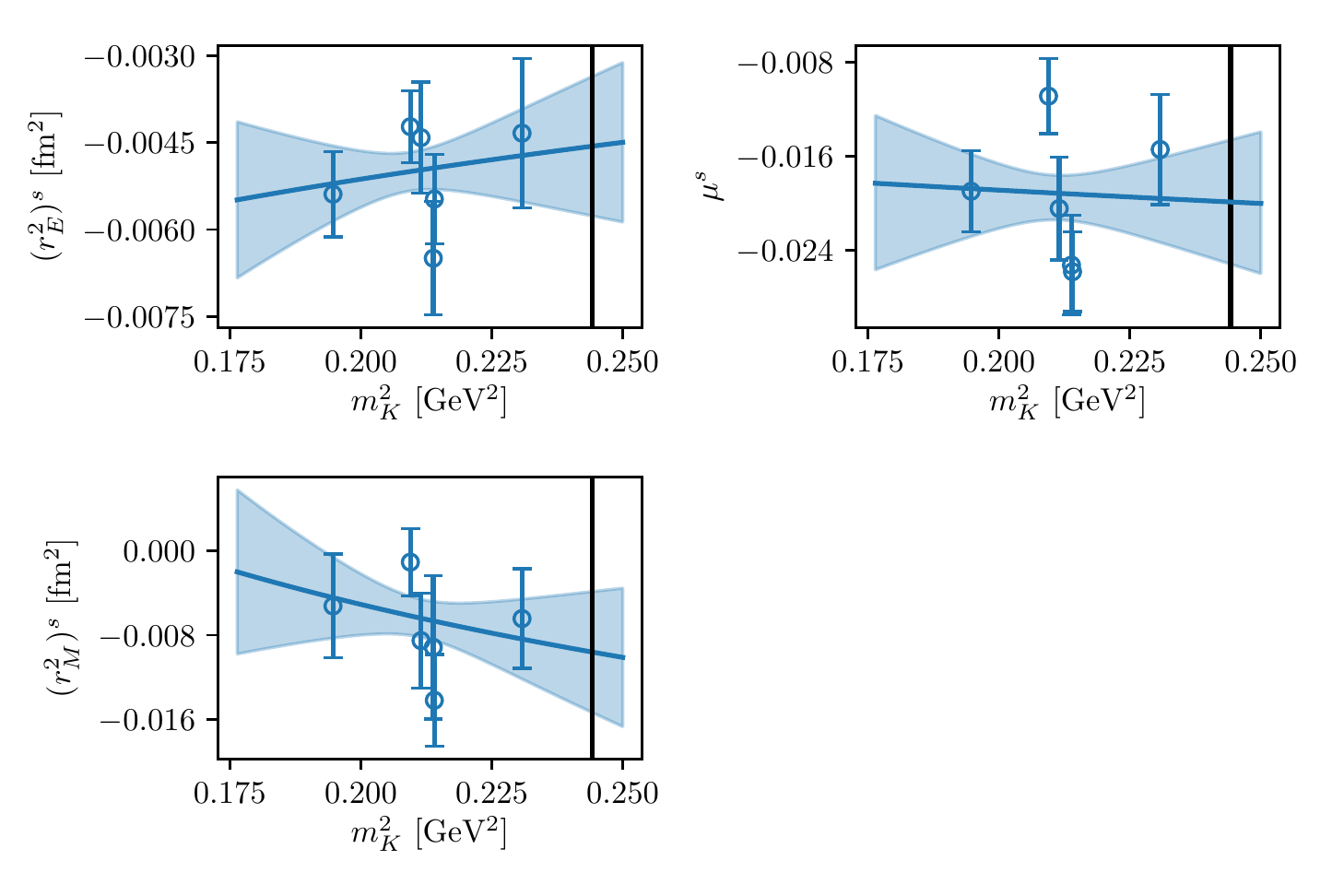}
\caption{Chiral and continuum extrapolation of the electric and magnetic radius and magnetic moment, using the standard method of Table~\ref{tab:chpt_cont_fit}. The vertical line denotes the physical kaon mass in the isospin limit \cite{Aoki:2016frl}.}
\label{fig:ccefit}
\end{figure}
From the
$z$-expansion fits, we can extract the strange magnetic moment $\mu^s$, as well as
the electric and magnetic charge radii
$(r^2_{E/M})^s$, 
\begin{align}
\mu^s &=\ a_0^M ,\\
(r^2_{E/M})^s &= -\ \dfrac{3}{2t_{\text{cut}}} a_1^{E/M} . \label{eq:radmag}
\end{align}
We have repeated the analysis in several variations in order to assess
systematic errors and subsequently perform chiral and continuum
extrapolations.  Since the radii and magnetic moments are defined at $Q^2=0$, we
perform the fits applying a cut of $Q^2<0.5$ GeV$^2$ and treat the difference
to fitting all of the data as a systematic uncertainty. This cut also ensures
that all ensembles contribute over the whole  range in $Q^2$.  
In total we thus have four sets of values for the radii and magnetic moments for
every ensemble, for which we analyze the lattice spacing and kaon mass
dependence.

The analyzed set of ensembles allow for a controlled chiral and continuum
extrapolation of the strange electromagnetic form factors.  In the following, we
will investigate the kaon mass dependence using
\begin{align}
( r^2_{E})^s(m_K)&=c_1+c_2 \log (m_K) +\tilde{c}_1   a^2 +c^L_1 \sqrt{L}e^{-m_K L}, \nonumber\\
\mu^s(m_K)&=c_3+c_4 m_K +\tilde{c}_2  a^2 + c^L_2 \Bigl(m_K-\frac{2}{L}\Bigr) e^{-m_K L}, \nonumber\\
(r^2_{M})^s(m_K)&=\frac{c_5}{m_K}+c_6+\tilde{c}_3 a^2 + c^L_3 \sqrt{L}e^{-m_K L} \label{eq:chipt_fit_form},
\end{align}
which is derived from  SU(3) heavy baryon chiral perturbation theory (HBChPT)
\cite{Hemmert:1999mr}, 
supplemented by terms describing the dependence on the lattice spacing
$a$ and the finite volume. (Note that the CLS ensembles
follow the $\tr M_q =$ constant trajectory, and so the kaon mass and the pion mass
are therefore not varied independently.) 
Since the finite-volume dependence originates exclusively from kaon loops, we
substitute the pion mass in the relevant expression for the magnetic moment
\cite{Beane:2004tw} by the mass of the kaon.
For a detailed discussion of the finite-volume dependence, we refer to  the
Supplemental Material \cite{supplmat:2019}.
	For the radii, we use the model-dependent ansatz of \cite{Sufian:2016pex,Sufian:2017osl}, assuming the 
	finite-volume dependence to be same as for the pion form factor calculated in
\cite{Tiburzi:2014yra}, again replacing the pion with the kaon mass.
Since our data for the magnetic radius do not show the divergent behavior
expected from HBChPT (see Fig.~\ref{fig:ccefit}), we amend the expressions from
\cite{Hemmert:1999mr} by the term $c_6$. While this cancellation of higher
order terms was already found in Ref.  ~\cite{Hammer:2002ei}, we note that the
convergence of HBChPT, the rate of which strongly depends on the observable, is,
in general, not easily established.

For each of the variations of the $z$-expansion fit in the previous section, we
analyze the chiral behavior separately.  The chirally extrapolated values for
the standard fit procedure and the variations of
the $z$-expansion fits performed to assess systematic uncertainties are given in
Table~\ref{tab:chpt_cont_fit}. We treat the difference of the central values for
the variations as an estimate for a (symmetric) systematic error. In addition,
we perform a fit including lattice artifacts or a fit including finite-volume dependence to the standard $z$-expansion fit. 
A simultaneous fit of the lattice spacing and finite-volume dependence amounts to the determination of four parameters from six data points for which the AIC$_c$ value is not defined. Therefore, we choose to perform separate extrapolations in our analysis.
The AIC$_c$ values, i.e., the Akaike information criterion \cite{tAKA73a} adjusted for small sample
size \cite{doi:10.1080/03610927808827599,10.1093/biomet/76.2.297}, for
the fits including lattice spacing or finite-volume effects, are larger by at least 24 in absolute value compared to the minimum AIC$_c$ (for the AIC$_c$ values, we use the maximum
likelihood estimator for the sample variance); i.e., the fits omitting
$\mathcal{O}(a^2),\mathcal{O}(\exp[-m_KL])$ are favored.
We therefore quote the fit
without lattice artifacts and finite-volume effects as our best value, using the difference in the
central value for the respective procedures as a systematic error from finite lattice
spacing and finite-volume corrections. 
\textbf{}
\begin{table}
\begin{tabular}{lccccc}
Fit & $(r_E^2)^s$ [fm$^2$] & $\mu^s$ & $(r_M^2)^s$ [fm$^2$]  & $\chi^2$/d.o.f.\\ \hline\hline
Standard & -0.0046(12) & -0.020(5) & -0.010(6) & 2.04(12) \\
Prior width &  -0.0053(15) & -0.020(6) & -0.012(8)& 1.47(12)\\
Plateau & -0.0045(14) & -0.022(8) &-0.014(8) & 1.62(12)\\
$\mathcal{O}(a^2)$ & -0.0036(16) &-0.009(7)& -0.003(8) & 1.91(9)  \\
$\mathcal{O}(\exp[-m_K L])$ & -0.0049(12) & -0.021(5) & -0.010(6) & 1.12(9)\\
No cut in $Q^2$ & -0.0051(9) & -0.017(5) & -0.008(5) & 3.14(12)
\end{tabular}
\caption{Fit results for the standard fit and variations thereof. 
}
\label{tab:chpt_cont_fit}
\end{table}
At the physical point, we find
\begin{align}
	(r^2_E)_{\text{phys}}^s &= -0.0046(12)(7)(1)(9)(3)(6)\,\text{fm}^2 ,\\
	\mu^s_\text{phys} &= -0.020(5)(0)(2)(11)(1)(3) ,\\
	(r^2_M)_{\text{phys}}^s &= -0.010(6)(2)(5)(7)(0)(2)\,\text{fm}^2 ,
\end{align}
as our final estimate, where the first error is statistical and the remaining
errors come from  the variations in the fitting procedure given in Table~\ref{tab:chpt_cont_fit}.

\begin{figure}[t]
\begin{center}
\includegraphics[scale=.9]{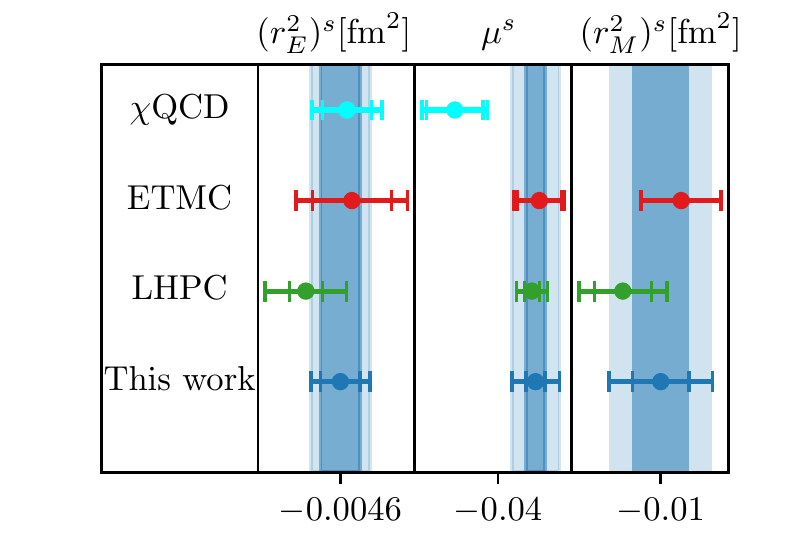}
\end{center}
\caption{Comparison of our final values for the radii and magnetic moments with
LHPC \cite{Green:2015wqa}, ETMC \cite{Alexandrou:2018zdf}, and $\chi$QCD
\cite{Sufian:2016pex,Sufian:2017osl}, where the dark and light blue bands
describe the  statistical error and the total error, including systematics,
respectively.}
\label{fig:comparison}
\end{figure}

\begin{figure}[b]
\includegraphics[scale=.6]{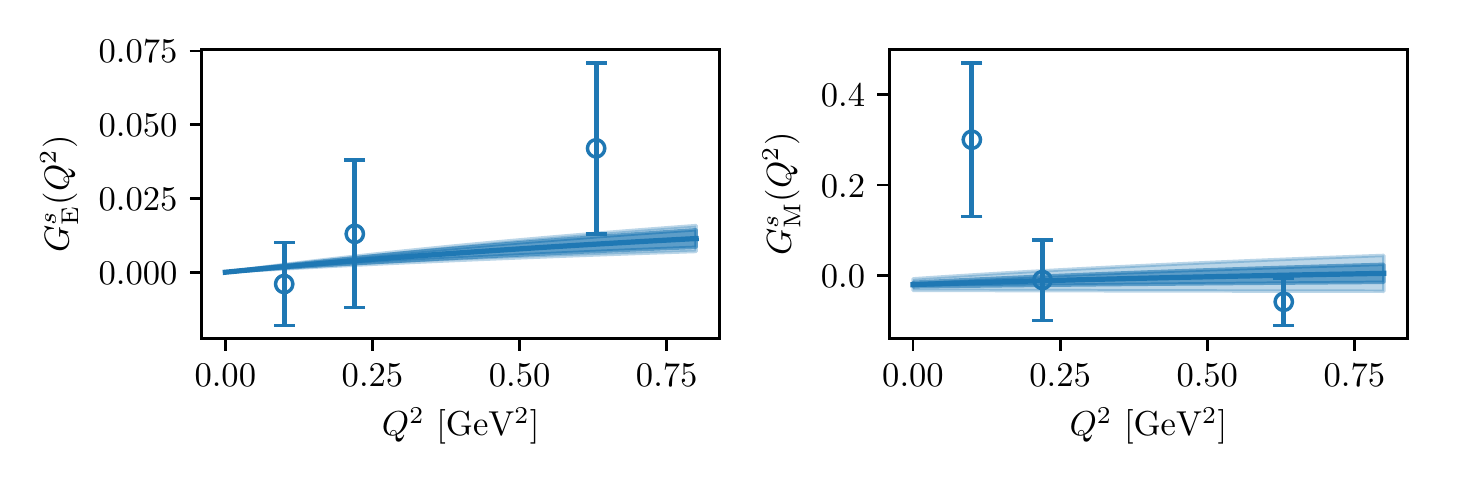}
\caption{Comparison of our standard fit, based on the $z$-expansion up to $k=1$, to the analysis of existing experimental data
\cite{Maas:2017snj}. The dark and light blue bands
describe the statistical error and the total error, including systematics,
respectively.}
\label{fig:comp_exp}
\end{figure}

For the radii, our values are in good agreement with other lattice
determinations \cite{Green:2015wqa,Alexandrou:2018zdf,Sufian:2016pex,Sufian:2017osl}. Our value
for 
the magnetic moment is again in good agreement with
\cite{Green:2015wqa,Alexandrou:2018zdf}. The magnetic moment from 
\cite{Sufian:2016pex,Sufian:2017osl} disagrees with our estimate and with
\cite{Green:2015wqa,Alexandrou:2018zdf} by more than 2 standard deviations,
see Fig.~\ref{fig:comparison}. 
Our best estimate of the radii and magnetic
moment compare favorably to the available experimental data, as can be seen from Fig.~\ref{fig:comp_exp}.

In summary, we have reported on our calculation of the strange contribution to the electromagnetic
form factors obtained on six CLS $N_f=2+1$ $\mathcal{O}(a)$-improved Wilson fermion
ensembles. For the calculation of the disconnected contributions, we use the
method of hierarchical probing, which significantly reduces the statistical error.
To deal with excited-state contamination, we employ the summation method. We find agreement with
plateau estimates for large enough source-sink separations. The strange charge
radii and the strange magnetic moment are obtained on each ensemble through
model independent $z$-expansion fits and later extrapolated to the physical
point. See the Supplemental Material \cite{supplmat:2019} for a
summary of the extracted form factors and $z$-expansion fits. Our results are
compatible with other lattice QCD studies and in good
agreement to experimental data. With the current set of ensembles, the physical values for the strange charge
radii and the strange magnetic moment still have large relative statistical
errors. We aim to improve this by enlarging the number of ensembles.

\begin{acknowledgments}
We thank H. Meyer, T. Harris, and G. von Hippel for useful discussions and comments.  This research is
supported by the Deutsche Forschungsgemeinschat (DFG, German Research Foundation) through the SFB 1044 ``The
low-energy frontier of the Standard Model". K.O. is supported by the DFG through Grant No. HI 2048/1-1.
Additionally, this work has been supported by the Cluster of Excellence
“Precision Physics, Fundamental Interactions, and Structure of Matter” (PRISMA$^+$
EXC 2118/1) funded by the German Research Foundation (DFG) within the German
Excellence Strategy (Project ID 39083149). Calculations for this project were
partly performed on the HPC clusters "{}Clover"{} and "{}HIMster II"{} at the
Helmholtz-Institut Mainz and "{}Mogon II"{} at JGU Mainz. Additional computer
time has been allocated through projects HMZ21 and HMZ36 on the BlueGene
supercomputer system "{}JUQUEEN"{} at NIC, J\"ulich. Our programs use the
QDP++ library \cite{Edwards:2004sx} and deflated SAP+GCR solver from the
openQCD package \cite{Luscher:2012av}, while the contractions have been
explicitly checked using \cite{Djukanovic:2016spv}. We are grateful to our
colleagues in the CLS initiative for sharing ensembles.
\end{acknowledgments}
\bibliography{lit}

\begin{thebibliography}{41}
\expandafter\ifx\csname natexlab\endcsname\relax\def\natexlab#1{#1}\fi
\expandafter\ifx\csname bibnamefont\endcsname\relax
  \def\bibnamefont#1{#1}\fi
\expandafter\ifx\csname bibfnamefont\endcsname\relax
  \def\bibfnamefont#1{#1}\fi
\expandafter\ifx\csname citenamefont\endcsname\relax
  \def\citenamefont#1{#1}\fi
\expandafter\ifx\csname url\endcsname\relax
  \def\url#1{\texttt{#1}}\fi
\expandafter\ifx\csname urlprefix\endcsname\relax\def\urlprefix{URL }\fi
\providecommand{\bibinfo}[2]{#2}
\providecommand{\eprint}[2][]{\url{#2}}

\bibitem[{\citenamefont{Spayde et~al.}(2004)}]{Spayde:2003nr}
\bibinfo{author}{\bibfnamefont{D.~T.} \bibnamefont{Spayde}}
  \bibnamefont{et~al.} (\bibinfo{collaboration}{SAMPLE}),
  \bibinfo{journal}{Phys. Lett.} \textbf{\bibinfo{volume}{B583}},
  \bibinfo{pages}{79} (\bibinfo{year}{2004}), \eprint{nucl-ex/0312016}.

\bibitem[{\citenamefont{Androi\ifmmode~\acute{c}\else \'{c}\fi{}
  et~al.}(2010)\citenamefont{Androi\ifmmode~\acute{c}\else \'{c}\fi{},
  Armstrong, Arvieux, Bailey, Beck, Beise, Benesch, Benmokhtar, Bimbot,
  Birchall et~al.}}]{PhysRevLett.104.012001}
\bibinfo{author}{\bibfnamefont{D.}~\bibnamefont{Androi\ifmmode~\acute{c}\else
  \'{c}\fi{}}}, \bibinfo{author}{\bibfnamefont{D.~S.} \bibnamefont{Armstrong}},
  \bibinfo{author}{\bibfnamefont{J.}~\bibnamefont{Arvieux}},
  \bibinfo{author}{\bibfnamefont{S.~L.} \bibnamefont{Bailey}},
  \bibinfo{author}{\bibfnamefont{D.~H.} \bibnamefont{Beck}},
  \bibinfo{author}{\bibfnamefont{E.~J.} \bibnamefont{Beise}},
  \bibinfo{author}{\bibfnamefont{J.}~\bibnamefont{Benesch}},
  \bibinfo{author}{\bibfnamefont{F.}~\bibnamefont{Benmokhtar}},
  \bibinfo{author}{\bibfnamefont{L.}~\bibnamefont{Bimbot}},
  \bibinfo{author}{\bibfnamefont{J.}~\bibnamefont{Birchall}},
  \bibnamefont{et~al.} (\bibinfo{collaboration}{G0 Collaboration}),
  \bibinfo{journal}{Phys. Rev. Lett.} \textbf{\bibinfo{volume}{104}},
  \bibinfo{pages}{012001} (\bibinfo{year}{2010}),
  \urlprefix\url{https://link.aps.org/doi/10.1103/PhysRevLett.104.012001}.

\bibitem[{\citenamefont{Armstrong et~al.}(2005)\citenamefont{Armstrong,
  Arvieux, Asaturyan, Averett, Bailey, Batigne, Beck, Beise, Benesch, Bimbot
  et~al.}}]{PhysRevLett.95.092001}
\bibinfo{author}{\bibfnamefont{D.~S.} \bibnamefont{Armstrong}},
  \bibinfo{author}{\bibfnamefont{J.}~\bibnamefont{Arvieux}},
  \bibinfo{author}{\bibfnamefont{R.}~\bibnamefont{Asaturyan}},
  \bibinfo{author}{\bibfnamefont{T.}~\bibnamefont{Averett}},
  \bibinfo{author}{\bibfnamefont{S.~L.} \bibnamefont{Bailey}},
  \bibinfo{author}{\bibfnamefont{G.}~\bibnamefont{Batigne}},
  \bibinfo{author}{\bibfnamefont{D.~H.} \bibnamefont{Beck}},
  \bibinfo{author}{\bibfnamefont{E.~J.} \bibnamefont{Beise}},
  \bibinfo{author}{\bibfnamefont{J.}~\bibnamefont{Benesch}},
  \bibinfo{author}{\bibfnamefont{L.}~\bibnamefont{Bimbot}},
  \bibnamefont{et~al.} (\bibinfo{collaboration}{G0 Collaboration}),
  \bibinfo{journal}{Phys. Rev. Lett.} \textbf{\bibinfo{volume}{95}},
  \bibinfo{pages}{092001} (\bibinfo{year}{2005}),
  \urlprefix\url{https://link.aps.org/doi/10.1103/PhysRevLett.95.092001}.

\bibitem[{\citenamefont{Baunack et~al.}(2009)\citenamefont{Baunack,
  Aulenbacher, Balaguer~R\'{\i}os, Capozza, Diefenbach, Gl\"aser, von Harrach,
  Imai, Kabu\ss{}, Kothe et~al.}}]{PhysRevLett.102.151803}
\bibinfo{author}{\bibfnamefont{S.}~\bibnamefont{Baunack}},
  \bibinfo{author}{\bibfnamefont{K.}~\bibnamefont{Aulenbacher}},
  \bibinfo{author}{\bibfnamefont{D.}~\bibnamefont{Balaguer~R\'{\i}os}},
  \bibinfo{author}{\bibfnamefont{L.}~\bibnamefont{Capozza}},
  \bibinfo{author}{\bibfnamefont{J.}~\bibnamefont{Diefenbach}},
  \bibinfo{author}{\bibfnamefont{B.}~\bibnamefont{Gl\"aser}},
  \bibinfo{author}{\bibfnamefont{D.}~\bibnamefont{von Harrach}},
  \bibinfo{author}{\bibfnamefont{Y.}~\bibnamefont{Imai}},
  \bibinfo{author}{\bibfnamefont{E.-M.} \bibnamefont{Kabu\ss{}}},
  \bibinfo{author}{\bibfnamefont{R.}~\bibnamefont{Kothe}},
  \bibnamefont{et~al.}, \bibinfo{journal}{Phys. Rev. Lett.}
  \textbf{\bibinfo{volume}{102}}, \bibinfo{pages}{151803}
  (\bibinfo{year}{2009}),
  \urlprefix\url{https://link.aps.org/doi/10.1103/PhysRevLett.102.151803}.

\bibitem[{\citenamefont{Ahmed et~al.}(2012)\citenamefont{Ahmed, Allada, Aniol,
  Armstrong, Arrington, Baturin, Bellini, Benesch, Beminiwattha, Benmokhtar
  et~al.}}]{PhysRevLett.108.102001}
\bibinfo{author}{\bibfnamefont{Z.}~\bibnamefont{Ahmed}},
  \bibinfo{author}{\bibfnamefont{K.}~\bibnamefont{Allada}},
  \bibinfo{author}{\bibfnamefont{K.~A.} \bibnamefont{Aniol}},
  \bibinfo{author}{\bibfnamefont{D.~S.} \bibnamefont{Armstrong}},
  \bibinfo{author}{\bibfnamefont{J.}~\bibnamefont{Arrington}},
  \bibinfo{author}{\bibfnamefont{P.}~\bibnamefont{Baturin}},
  \bibinfo{author}{\bibfnamefont{V.}~\bibnamefont{Bellini}},
  \bibinfo{author}{\bibfnamefont{J.}~\bibnamefont{Benesch}},
  \bibinfo{author}{\bibfnamefont{R.}~\bibnamefont{Beminiwattha}},
  \bibinfo{author}{\bibfnamefont{F.}~\bibnamefont{Benmokhtar}},
  \bibnamefont{et~al.} (\bibinfo{collaboration}{HAPPEX Collaboration}),
  \bibinfo{journal}{Phys. Rev. Lett.} \textbf{\bibinfo{volume}{108}},
  \bibinfo{pages}{102001} (\bibinfo{year}{2012}),
  \urlprefix\url{https://link.aps.org/doi/10.1103/PhysRevLett.108.102001}.

\bibitem[{\citenamefont{Aniol et~al.}(2004)}]{Aniol:2004hp}
\bibinfo{author}{\bibfnamefont{K.~A.} \bibnamefont{Aniol}} \bibnamefont{et~al.}
  (\bibinfo{collaboration}{HAPPEX}), \bibinfo{journal}{Phys. Rev.}
  \textbf{\bibinfo{volume}{C69}}, \bibinfo{pages}{065501}
  (\bibinfo{year}{2004}), \eprint{nucl-ex/0402004}.

\bibitem[{\citenamefont{Maas and Paschke}(2017)}]{Maas:2017snj}
\bibinfo{author}{\bibfnamefont{F.~E.} \bibnamefont{Maas}} \bibnamefont{and}
  \bibinfo{author}{\bibfnamefont{K.~D.} \bibnamefont{Paschke}},
  \bibinfo{journal}{Prog. Part. Nucl. Phys.} \textbf{\bibinfo{volume}{95}},
  \bibinfo{pages}{209} (\bibinfo{year}{2017}).

\bibitem[{\citenamefont{Stathopoulos et~al.}(2013)\citenamefont{Stathopoulos,
  Laeuchli, and Orginos}}]{Stathopoulos:2013aci}
\bibinfo{author}{\bibfnamefont{A.}~\bibnamefont{Stathopoulos}},
  \bibinfo{author}{\bibfnamefont{J.}~\bibnamefont{Laeuchli}}, \bibnamefont{and}
  \bibinfo{author}{\bibfnamefont{K.}~\bibnamefont{Orginos}}
  (\bibinfo{year}{2013}), \eprint{1302.4018}.

\bibitem[{\citenamefont{Becker et~al.}(2018)}]{Becker:2018ggl}
\bibinfo{author}{\bibfnamefont{D.}~\bibnamefont{Becker}} \bibnamefont{et~al.},
  \bibinfo{journal}{Eur. Phys. J.} \textbf{\bibinfo{volume}{A54}},
  \bibinfo{pages}{208} (\bibinfo{year}{2018}), \eprint{1802.04759}.

\bibitem[{\citenamefont{Djukanovic et~al.}(2018)\citenamefont{Djukanovic,
  Meyer, Ottnad, von Hippel, Wilhelm, and Wittig}}]{Djukanovic:2018iir}
\bibinfo{author}{\bibfnamefont{D.}~\bibnamefont{Djukanovic}},
  \bibinfo{author}{\bibfnamefont{H.}~\bibnamefont{Meyer}},
  \bibinfo{author}{\bibfnamefont{K.}~\bibnamefont{Ottnad}},
  \bibinfo{author}{\bibfnamefont{G.}~\bibnamefont{von Hippel}},
  \bibinfo{author}{\bibfnamefont{J.}~\bibnamefont{Wilhelm}}, \bibnamefont{and}
  \bibinfo{author}{\bibfnamefont{H.}~\bibnamefont{Wittig}}, in
  \emph{\bibinfo{booktitle}{{36th International Symposium on Lattice Field
  Theory (Lattice 2018) East Lansing, MI, United States, July 22-28, 2018}}}
  (\bibinfo{year}{2018}), \eprint{1810.10810}.

\bibitem[{\citenamefont{Bruno et~al.}(2015)\citenamefont{Bruno, Djukanovic,
  Engel, Francis, Herdoiza, Horch, Korcyl, Korzec, Papinutto, Schaefer
  et~al.}}]{Bruno2015}
\bibinfo{author}{\bibfnamefont{M.}~\bibnamefont{Bruno}},
  \bibinfo{author}{\bibfnamefont{D.}~\bibnamefont{Djukanovic}},
  \bibinfo{author}{\bibfnamefont{G.~P.} \bibnamefont{Engel}},
  \bibinfo{author}{\bibfnamefont{A.}~\bibnamefont{Francis}},
  \bibinfo{author}{\bibfnamefont{G.}~\bibnamefont{Herdoiza}},
  \bibinfo{author}{\bibfnamefont{H.}~\bibnamefont{Horch}},
  \bibinfo{author}{\bibfnamefont{P.}~\bibnamefont{Korcyl}},
  \bibinfo{author}{\bibfnamefont{T.}~\bibnamefont{Korzec}},
  \bibinfo{author}{\bibfnamefont{M.}~\bibnamefont{Papinutto}},
  \bibinfo{author}{\bibfnamefont{S.}~\bibnamefont{Schaefer}},
  \bibnamefont{et~al.}, \bibinfo{journal}{Journal of High Energy Physics}
  \textbf{\bibinfo{volume}{2015}}, \bibinfo{pages}{43} (\bibinfo{year}{2015}),
  ISSN \bibinfo{issn}{1029-8479},
  \urlprefix\url{https://doi.org/10.1007/JHEP02(2015)043}.

\bibitem[{\citenamefont{Luscher and Schaefer}(2011)}]{Luscher:2011kk}
\bibinfo{author}{\bibfnamefont{M.}~\bibnamefont{L\"uscher}} \bibnamefont{and}
  \bibinfo{author}{\bibfnamefont{S.}~\bibnamefont{Schaefer}},
  \bibinfo{journal}{JHEP} \textbf{\bibinfo{volume}{07}}, \bibinfo{pages}{036}
  (\bibinfo{year}{2011}), \eprint{1105.4749}.

\bibitem[{\citenamefont{Bietenholz et~al.}(2010)\citenamefont{Bietenholz,
  Bornyakov, Cundy, G{\"o}ckeler, Horsley, Kennedy, Lockhart, Nakamura, Perlt,
  Pleiter et~al.}}]{BIETENHOLZ2010436}
\bibinfo{author}{\bibfnamefont{W.}~\bibnamefont{Bietenholz}},
  \bibinfo{author}{\bibfnamefont{V.}~\bibnamefont{Bornyakov}},
  \bibinfo{author}{\bibfnamefont{N.}~\bibnamefont{Cundy}},
  \bibinfo{author}{\bibfnamefont{M.}~\bibnamefont{G{\"o}ckeler}},
  \bibinfo{author}{\bibfnamefont{R.}~\bibnamefont{Horsley}},
  \bibinfo{author}{\bibfnamefont{A.}~\bibnamefont{Kennedy}},
  \bibinfo{author}{\bibfnamefont{W.}~\bibnamefont{Lockhart}},
  \bibinfo{author}{\bibfnamefont{Y.}~\bibnamefont{Nakamura}},
  \bibinfo{author}{\bibfnamefont{H.}~\bibnamefont{Perlt}},
  \bibinfo{author}{\bibfnamefont{D.}~\bibnamefont{Pleiter}},
  \bibnamefont{et~al.}, \bibinfo{journal}{Physics Letters B}
  \textbf{\bibinfo{volume}{690}}, \bibinfo{pages}{436 } (\bibinfo{year}{2010}),
  ISSN \bibinfo{issn}{0370-2693},
  \urlprefix\url{http://www.sciencedirect.com/science/article/pii/S0370269310006702}.

\bibitem[{\citenamefont{Bruno et~al.}(2017)\citenamefont{Bruno, Korzec, and
  Schaefer}}]{Bruno:2016plf}
\bibinfo{author}{\bibfnamefont{M.}~\bibnamefont{Bruno}},
  \bibinfo{author}{\bibfnamefont{T.}~\bibnamefont{Korzec}}, \bibnamefont{and}
  \bibinfo{author}{\bibfnamefont{S.}~\bibnamefont{Schaefer}},
  \bibinfo{journal}{Phys. Rev.} \textbf{\bibinfo{volume}{D95}},
  \bibinfo{pages}{074504} (\bibinfo{year}{2017}), \eprint{1608.08900}.

\bibitem[{\citenamefont{C\`{e} et~al.}(2018)\citenamefont{C\`{e}, G\'{e}rardin, Ottnad,
  and Meyer}}]{Ce:2018ziv}
\bibinfo{author}{\bibfnamefont{M.}~\bibnamefont{C\`{e}}},
  \bibinfo{author}{\bibfnamefont{A.}~\bibnamefont{G\'{e}rardin}},
  \bibinfo{author}{\bibfnamefont{K.}~\bibnamefont{Ottnad}}, \bibnamefont{and}
  \bibinfo{author}{\bibfnamefont{H.~B.} \bibnamefont{Meyer}},
  \bibinfo{journal}{PoS} \textbf{\bibinfo{volume}{LATTICE2018}},
  \bibinfo{pages}{137} (\bibinfo{year}{2018}), \eprint{1811.08669}.

\bibitem[{\citenamefont{Gusken}(1990)}]{Gusken:1989qx}
\bibinfo{author}{\bibfnamefont{S.}~\bibnamefont{G\"usken}},
  \bibinfo{journal}{Nucl. Phys. Proc. Suppl.} \textbf{\bibinfo{volume}{17}},
  \bibinfo{pages}{361} (\bibinfo{year}{1990}).

\bibitem[{\citenamefont{Bali et~al.}(2015)\citenamefont{Bali, Collins, Frommer,
  Kahl, Kanamori, M{\"u}ller, Rottmann, and Simeth}}]{Bali:2015qya}
\bibinfo{author}{\bibfnamefont{G.}~\bibnamefont{Bali}},
  \bibinfo{author}{\bibfnamefont{S.}~\bibnamefont{Collins}},
  \bibinfo{author}{\bibfnamefont{A.}~\bibnamefont{Frommer}},
  \bibinfo{author}{\bibfnamefont{K.}~\bibnamefont{Kahl}},
  \bibinfo{author}{\bibfnamefont{I.}~\bibnamefont{Kanamori}},
  \bibinfo{author}{\bibfnamefont{B.}~\bibnamefont{M{\"u}ller}},
  \bibinfo{author}{\bibfnamefont{M.}~\bibnamefont{Rottmann}}, \bibnamefont{and}
  \bibinfo{author}{\bibfnamefont{J.}~\bibnamefont{Simeth}},
  \bibinfo{journal}{PoS} \textbf{\bibinfo{volume}{LATTICE2015}},
  \bibinfo{pages}{350} (\bibinfo{year}{2015}), \eprint{1509.06865}.

\bibitem[{\citenamefont{Shintani et~al.}(2015)\citenamefont{Shintani, Arthur,
  Blum, Izubuchi, Jung, and Lehner}}]{PhysRevD.91.114511}
\bibinfo{author}{\bibfnamefont{E.}~\bibnamefont{Shintani}},
  \bibinfo{author}{\bibfnamefont{R.}~\bibnamefont{Arthur}},
  \bibinfo{author}{\bibfnamefont{T.}~\bibnamefont{Blum}},
  \bibinfo{author}{\bibfnamefont{T.}~\bibnamefont{Izubuchi}},
  \bibinfo{author}{\bibfnamefont{C.}~\bibnamefont{Jung}}, \bibnamefont{and}
  \bibinfo{author}{\bibfnamefont{C.}~\bibnamefont{Lehner}},
  \bibinfo{journal}{Phys. Rev. D} \textbf{\bibinfo{volume}{91}},
  \bibinfo{pages}{114511} (\bibinfo{year}{2015}),
  \urlprefix\url{https://link.aps.org/doi/10.1103/PhysRevD.91.114511}.

\bibitem[{\citenamefont{G\'erardin et~al.}(2019)\citenamefont{G\'erardin,
  Harris, and Meyer}}]{PhysRevD.99.014519}
\bibinfo{author}{\bibfnamefont{A.}~\bibnamefont{G\'erardin}},
  \bibinfo{author}{\bibfnamefont{T.}~\bibnamefont{Harris}}, \bibnamefont{and}
  \bibinfo{author}{\bibfnamefont{H.~B.} \bibnamefont{Meyer}},
  \bibinfo{journal}{Phys. Rev. D} \textbf{\bibinfo{volume}{99}},
  \bibinfo{pages}{014519} (\bibinfo{year}{2019}),
  \urlprefix\url{https://link.aps.org/doi/10.1103/PhysRevD.99.014519}.

\bibitem[{\citenamefont{Alexandrou et~al.}(2008)\citenamefont{Alexandrou,
  Korzec, Koutsou, Brinet, Carbonell, Drach, Harraud, and
  Baron}}]{Alexandrou:2008rp}
\bibinfo{author}{\bibfnamefont{C.}~\bibnamefont{Alexandrou}},
  \bibinfo{author}{\bibfnamefont{T.}~\bibnamefont{Korzec}},
  \bibinfo{author}{\bibfnamefont{G.}~\bibnamefont{Koutsou}},
  \bibinfo{author}{\bibfnamefont{M.}~\bibnamefont{Brinet}},
  \bibinfo{author}{\bibfnamefont{J.}~\bibnamefont{Carbonell}},
  \bibinfo{author}{\bibfnamefont{V.}~\bibnamefont{Drach}},
  \bibinfo{author}{\bibfnamefont{P.-A.} \bibnamefont{Harraud}},
  \bibnamefont{and} \bibinfo{author}{\bibfnamefont{R.}~\bibnamefont{Baron}}
  (\bibinfo{collaboration}{European Twisted Mass}), \bibinfo{journal}{PoS}
  \textbf{\bibinfo{volume}{LATTICE2008}}, \bibinfo{pages}{139}
  (\bibinfo{year}{2008}), \eprint{0811.0724}.

\bibitem[{\citenamefont{Green et~al.}(2014)\citenamefont{Green, Negele,
  Pochinsky, Syritsyn, Engelhardt, and Krieg}}]{Green:2014xba}
\bibinfo{author}{\bibfnamefont{J.~R.} \bibnamefont{Green}},
  \bibinfo{author}{\bibfnamefont{J.~W.} \bibnamefont{Negele}},
  \bibinfo{author}{\bibfnamefont{A.~V.} \bibnamefont{Pochinsky}},
  \bibinfo{author}{\bibfnamefont{S.~N.} \bibnamefont{Syritsyn}},
  \bibinfo{author}{\bibfnamefont{M.}~\bibnamefont{Engelhardt}},
  \bibnamefont{and} \bibinfo{author}{\bibfnamefont{S.}~\bibnamefont{Krieg}},
  \bibinfo{journal}{Phys. Rev.} \textbf{\bibinfo{volume}{D90}},
  \bibinfo{pages}{074507} (\bibinfo{year}{2014}), \eprint{1404.4029}.

\bibitem[{\citenamefont{Capitani et~al.}(2015)\citenamefont{Capitani,
  Della~Morte, Djukanovic, von Hippel, Hua, Jäger, Knippschild, Meyer, Rae,
  and Wittig}}]{Capitani:2015sba}
\bibinfo{author}{\bibfnamefont{S.}~\bibnamefont{Capitani}},
  \bibinfo{author}{\bibfnamefont{M.}~\bibnamefont{Della~Morte}},
  \bibinfo{author}{\bibfnamefont{D.}~\bibnamefont{Djukanovic}},
  \bibinfo{author}{\bibfnamefont{G.}~\bibnamefont{von Hippel}},
  \bibinfo{author}{\bibfnamefont{J.}~\bibnamefont{Hua}},
  \bibinfo{author}{\bibfnamefont{B.}~\bibnamefont{J\"ager}},
  \bibinfo{author}{\bibfnamefont{B.}~\bibnamefont{Knippschild}},
  \bibinfo{author}{\bibfnamefont{H.~B.} \bibnamefont{Meyer}},
  \bibinfo{author}{\bibfnamefont{T.~D.} \bibnamefont{Rae}}, \bibnamefont{and}
  \bibinfo{author}{\bibfnamefont{H.}~\bibnamefont{Wittig}},
  \bibinfo{journal}{Phys. Rev.} \textbf{\bibinfo{volume}{D92}},
  \bibinfo{pages}{054511} (\bibinfo{year}{2015}), \eprint{1504.04628}.

\bibitem[{\citenamefont{Capitani et~al.}(2019)\citenamefont{Capitani,
  Della~Morte, Djukanovic, von Hippel, Hua, J{\"a}ger, Junnarkar, Meyer, Rae,
  and Wittig}}]{Capitani:2017qpc}
\bibinfo{author}{\bibfnamefont{S.}~\bibnamefont{Capitani}},
  \bibinfo{author}{\bibfnamefont{M.}~\bibnamefont{Della~Morte}},
  \bibinfo{author}{\bibfnamefont{D.}~\bibnamefont{Djukanovic}},
  \bibinfo{author}{\bibfnamefont{G.~M.} \bibnamefont{von Hippel}},
  \bibinfo{author}{\bibfnamefont{J.}~\bibnamefont{Hua}},
  \bibinfo{author}{\bibfnamefont{B.}~\bibnamefont{J{\"a}ger}},
  \bibinfo{author}{\bibfnamefont{P.~M.} \bibnamefont{Junnarkar}},
  \bibinfo{author}{\bibfnamefont{H.~B.} \bibnamefont{Meyer}},
  \bibinfo{author}{\bibfnamefont{T.~D.} \bibnamefont{Rae}}, \bibnamefont{and}
  \bibinfo{author}{\bibfnamefont{H.}~\bibnamefont{Wittig}},
  \bibinfo{journal}{Int. J. Mod. Phys.} \textbf{\bibinfo{volume}{A34}},
  \bibinfo{pages}{1950009} (\bibinfo{year}{2019}), \eprint{1705.06186},
  \urlprefix\url{https://doi.org/10.1142/S0217751X1950009X}.

\bibitem[{\citenamefont{Syritsyn et~al.}(2010)\citenamefont{Syritsyn, Bratt,
  Lin, Meyer, Negele, Pochinsky, Procura, Engelhardt, H\"agler, Hemmert
  et~al.}}]{PhysRevD.81.034507}
\bibinfo{author}{\bibfnamefont{S.~N.} \bibnamefont{Syritsyn}},
  \bibinfo{author}{\bibfnamefont{J.~D.} \bibnamefont{Bratt}},
  \bibinfo{author}{\bibfnamefont{M.~F.} \bibnamefont{Lin}},
  \bibinfo{author}{\bibfnamefont{H.~B.} \bibnamefont{Meyer}},
  \bibinfo{author}{\bibfnamefont{J.~W.} \bibnamefont{Negele}},
  \bibinfo{author}{\bibfnamefont{A.~V.} \bibnamefont{Pochinsky}},
  \bibinfo{author}{\bibfnamefont{M.}~\bibnamefont{Procura}},
  \bibinfo{author}{\bibfnamefont{M.}~\bibnamefont{Engelhardt}},
  \bibinfo{author}{\bibfnamefont{P.}~\bibnamefont{H\"agler}},
  \bibinfo{author}{\bibfnamefont{T.~R.} \bibnamefont{Hemmert}},
  \bibnamefont{et~al.} (\bibinfo{collaboration}{LHPC Collaboration}),
  \bibinfo{journal}{Phys. Rev. D} \textbf{\bibinfo{volume}{81}},
  \bibinfo{pages}{034507} (\bibinfo{year}{2010}),
  \urlprefix\url{https://link.aps.org/doi/10.1103/PhysRevD.81.034507}.

\bibitem[{\citenamefont{Maiani et~al.}(1987)\citenamefont{Maiani, Martinelli,
  Paciello, and Taglienti}}]{Maiani:1987by}
\bibinfo{author}{\bibfnamefont{L.}~\bibnamefont{Maiani}},
  \bibinfo{author}{\bibfnamefont{G.}~\bibnamefont{Martinelli}},
  \bibinfo{author}{\bibfnamefont{M.~L.} \bibnamefont{Paciello}},
  \bibnamefont{and}
  \bibinfo{author}{\bibfnamefont{B.}~\bibnamefont{Taglienti}},
  \bibinfo{journal}{Nucl. Phys.} \textbf{\bibinfo{volume}{B293}},
  \bibinfo{pages}{420} (\bibinfo{year}{1987}).

\bibitem[{\citenamefont{Doi et~al.}(2009)\citenamefont{Doi, Deka, Dong, Draper,
  Liu, Mankame, Mathur, and Streuer}}]{Doi:2009sq}
\bibinfo{author}{\bibfnamefont{T.}~\bibnamefont{Doi}},
  \bibinfo{author}{\bibfnamefont{M.}~\bibnamefont{Deka}},
  \bibinfo{author}{\bibfnamefont{S.-J.} \bibnamefont{Dong}},
  \bibinfo{author}{\bibfnamefont{T.}~\bibnamefont{Draper}},
  \bibinfo{author}{\bibfnamefont{K.-F.} \bibnamefont{Liu}},
  \bibinfo{author}{\bibfnamefont{D.}~\bibnamefont{Mankame}},
  \bibinfo{author}{\bibfnamefont{N.}~\bibnamefont{Mathur}}, \bibnamefont{and}
  \bibinfo{author}{\bibfnamefont{T.}~\bibnamefont{Streuer}},
  \bibinfo{journal}{Phys. Rev.} \textbf{\bibinfo{volume}{D80}},
  \bibinfo{pages}{094503} (\bibinfo{year}{2009}), \eprint{0903.3232}.

\bibitem[{\citenamefont{Brandt et~al.}(2011)\citenamefont{Brandt, Capitani,
  Della~Morte, Djukanovic, Gegelia, von Hippel, Juttner, Knippschild, Meyer,
  and Wittig}}]{Brandt:2011sj}
\bibinfo{author}{\bibfnamefont{B.~B.} \bibnamefont{Brandt}},
  \bibinfo{author}{\bibfnamefont{S.}~\bibnamefont{Capitani}},
  \bibinfo{author}{\bibfnamefont{M.}~\bibnamefont{Della~Morte}},
  \bibinfo{author}{\bibfnamefont{D.}~\bibnamefont{Djukanovic}},
  \bibinfo{author}{\bibfnamefont{J.}~\bibnamefont{Gegelia}},
  \bibinfo{author}{\bibfnamefont{G.}~\bibnamefont{von Hippel}},
  \bibinfo{author}{\bibfnamefont{A.}~\bibnamefont{J\"uttner}},
  \bibinfo{author}{\bibfnamefont{B.}~\bibnamefont{Knippschild}},
  \bibinfo{author}{\bibfnamefont{H.~B.} \bibnamefont{Meyer}}, \bibnamefont{and}
  \bibinfo{author}{\bibfnamefont{H.}~\bibnamefont{Wittig}},
  \bibinfo{journal}{Eur. Phys. J. ST} \textbf{\bibinfo{volume}{198}},
  \bibinfo{pages}{79} (\bibinfo{year}{2011}), \eprint{1106.1554}.

\bibitem[{\citenamefont{ Supplemental Material at [URL will be inserted by publisher]}(2019)}]{supplmat:2019}
	See supplemental Material in the appendix, for a
	detailed discussion of the finite volume effects, which includes Ref.~\cite{Bernard:1995dp}, the
	effective mass plot of the nucleon for the ensemble N200, and summary
	tables of the extracted form factors and z-expansion fits for every
	ensemble.

\bibitem{Bernard:1995dp}
	\bibinfo{author}{\bibfnamefont{V.}~\bibnamefont{Bernard}}, 
	\bibinfo{author}{\bibfnamefont{N.}~\bibnamefont{Kaiser}}, \bibnamefont{and}
	\bibinfo{author}{\bibfnamefont{U.-G.} \bibnamefont{Mei{\ss}ner}}, 
	\bibinfo{journal}{Int. J. Mod. Phys.} \textbf{\bibinfo{volume}{E4}},
	\bibinfo{pages}{193--346} (\bibinfo{year}{1995}).

\bibitem[{\citenamefont{Hill and Paz}(2010)}]{PhysRevD.82.113005}
\bibinfo{author}{\bibfnamefont{R.~J.} \bibnamefont{Hill}} \bibnamefont{and}
  \bibinfo{author}{\bibfnamefont{G.}~\bibnamefont{Paz}},
  \bibinfo{journal}{Phys. Rev. D} \textbf{\bibinfo{volume}{82}},
  \bibinfo{pages}{113005} (\bibinfo{year}{2010}),
  \urlprefix\url{https://link.aps.org/doi/10.1103/PhysRevD.82.113005}.

\bibitem[{\citenamefont{Epstein et~al.}(2014)\citenamefont{Epstein, Paz, and
  Roy}}]{PhysRevD.90.074027}
\bibinfo{author}{\bibfnamefont{Z.}~\bibnamefont{Epstein}},
  \bibinfo{author}{\bibfnamefont{G.}~\bibnamefont{Paz}}, \bibnamefont{and}
  \bibinfo{author}{\bibfnamefont{J.}~\bibnamefont{Roy}},
  \bibinfo{journal}{Phys. Rev. D} \textbf{\bibinfo{volume}{90}},
  \bibinfo{pages}{074027} (\bibinfo{year}{2014}),
  \urlprefix\url{https://link.aps.org/doi/10.1103/PhysRevD.90.074027}.

\bibitem[{\citenamefont{Hemmert et~al.}(1999)\citenamefont{Hemmert, Kubis, and
  Meissner}}]{Hemmert:1999mr}
\bibinfo{author}{\bibfnamefont{T.~R.} \bibnamefont{Hemmert}},
  \bibinfo{author}{\bibfnamefont{B.}~\bibnamefont{Kubis}}, \bibnamefont{and}
  \bibinfo{author}{\bibfnamefont{U.~G.} \bibnamefont{Mei\ss ner}},
  \bibinfo{journal}{Phys. Rev.} \textbf{\bibinfo{volume}{C60}},
  \bibinfo{pages}{045501} (\bibinfo{year}{1999}), \eprint{nucl-th/9904076}.

\bibitem[{\citenamefont{Beane}(2004)}]{Beane:2004tw}
\bibinfo{author}{\bibfnamefont{S.~R.} \bibnamefont{Beane}},
  \bibinfo{journal}{Phys. Rev.} \textbf{\bibinfo{volume}{D70}},
  \bibinfo{pages}{034507} (\bibinfo{year}{2004}), \eprint{hep-lat/0403015}.


\bibitem[{\citenamefont{Sufian et~al.}(2017)\citenamefont{Sufian, Yang, Liang,
  Draper, and Liu}}]{Sufian:2016pex}
\bibinfo{author}{\bibfnamefont{R.~S.} \bibnamefont{Sufian}},
  \bibinfo{author}{\bibfnamefont{Y.-B.} \bibnamefont{Yang}},
  \bibinfo{author}{\bibfnamefont{A.} \bibnamefont{Alexandru}},
  \bibinfo{author}{\bibfnamefont{T.}~\bibnamefont{Draper}}, 
  \bibinfo{author}{\bibfnamefont{J.}~\bibnamefont{Liang}}, \bibnamefont{and}
  \bibinfo{author}{\bibfnamefont{K.-F.} \bibnamefont{Liu}},
  \bibinfo{journal}{Phys. Rev. Lett.} \textbf{\bibinfo{volume}{118}},
  \bibinfo{pages}{042001} (\bibinfo{year}{2017}), \eprint{1606.07075}.

\bibitem[{\citenamefont{Sufian et~al.}(2017)\citenamefont{Sufian, Yang, Liang,
  Draper, and Liu}}]{Sufian:2017osl}
\bibinfo{author}{\bibfnamefont{R.~S.} \bibnamefont{Sufian}},
  \bibinfo{author}{\bibfnamefont{Y.-B.} \bibnamefont{Yang}},
  \bibinfo{author}{\bibfnamefont{J.}~\bibnamefont{Liang}},
  \bibinfo{author}{\bibfnamefont{T.}~\bibnamefont{Draper}}, \bibnamefont{and}
  \bibinfo{author}{\bibfnamefont{K.-F.} \bibnamefont{Liu}},
  \bibinfo{journal}{Phys. Rev.} \textbf{\bibinfo{volume}{D96}},
  \bibinfo{pages}{114504} (\bibinfo{year}{2017}), \eprint{1705.05849}.
%

\bibitem[{\citenamefont{Tiburzi}(2014)}]{Tiburzi:2014yra}
\bibinfo{author}{\bibfnamefont{B.~C.} \bibnamefont{Tiburzi}},
  \bibinfo{journal}{Phys. Rev.} \textbf{\bibinfo{volume}{D90}},
  \bibinfo{pages}{054508} (\bibinfo{year}{2014}), \eprint{1407.4059}.

\bibitem[{\citenamefont{Hammer et~al.}(2003)\citenamefont{Hammer, Puglia,
  Ramsey-Musolf, and Zhu}}]{Hammer:2002ei}
\bibinfo{author}{\bibfnamefont{H.~W.} \bibnamefont{Hammer}},
  \bibinfo{author}{\bibfnamefont{S.~J.} \bibnamefont{Puglia}},
  \bibinfo{author}{\bibfnamefont{M.~J.} \bibnamefont{Ramsey-Musolf}},
  \bibnamefont{and} \bibinfo{author}{\bibfnamefont{S.-L.} \bibnamefont{Zhu}},
  \bibinfo{journal}{Phys. Lett.} \textbf{\bibinfo{volume}{B562}},
  \bibinfo{pages}{208} (\bibinfo{year}{2003}), \eprint{hep-ph/0206301}.

\bibitem[{\citenamefont{Akaike}(1973)}]{tAKA73a}
\bibinfo{author}{\bibfnamefont{H.}~\bibnamefont{Akaike}}, in
  \emph{\bibinfo{booktitle}{Proc. 2nd International Symposium on Information
  Theory,(Eds. B. N. Petrov and F. Csaki), Akademiai Kiado, Budapest}}
  (\bibinfo{year}{1973}), pp. \bibinfo{pages}{267--281}.

\bibitem[{\citenamefont{Sugiura}(1978)}]{doi:10.1080/03610927808827599}
\bibinfo{author}{\bibfnamefont{N.}~\bibnamefont{Sugiura}},
  \bibinfo{journal}{Communications in Statistics - Theory and Methods}
  \textbf{\bibinfo{volume}{7}}, \bibinfo{pages}{13} (\bibinfo{year}{1978}),
  \urlprefix\url{https://doi.org/10.1080/03610927808827599}.

\bibitem[{\citenamefont{HURVICH and TSAI}(1989)}]{10.1093/biomet/76.2.297}
\bibinfo{author}{\bibfnamefont{C.~M.} \bibnamefont{Hurvich}} \bibnamefont{and}
  \bibinfo{author}{\bibfnamefont{C.-L.} \bibnamefont{Tsai}},
  \bibinfo{journal}{Biometrika} \textbf{\bibinfo{volume}{76}},
  \bibinfo{pages}{297} (\bibinfo{year}{1989}), ISSN \bibinfo{issn}{0006-3444},
  \eprint{http://oup.prod.sis.lan/biomet/article-pdf/76/2/297/737009/76-2-297.pdf},
  \urlprefix\url{https://dx.doi.org/10.1093/biomet/76.2.297}.

\bibitem[{\citenamefont{Aoki et~al.}(2017)}]{Aoki:2016frl}
\bibinfo{author}{\bibfnamefont{S.}~\bibnamefont{Aoki}} \bibnamefont{et~al.},
  \bibinfo{journal}{Eur. Phys. J.} \textbf{\bibinfo{volume}{C77}},
  \bibinfo{pages}{112} (\bibinfo{year}{2017}), \eprint{1607.00299}.

\bibitem[{\citenamefont{Green et~al.}(2015)\citenamefont{Green, Meinel,
  Engelhardt, Krieg, Laeuchli, Negele, Orginos, Pochinsky, and
  Syritsyn}}]{Green:2015wqa}
\bibinfo{author}{\bibfnamefont{J.}~\bibnamefont{Green}},
  \bibinfo{author}{\bibfnamefont{S.}~\bibnamefont{Meinel}},
  \bibinfo{author}{\bibfnamefont{M.}~\bibnamefont{Engelhardt}},
  \bibinfo{author}{\bibfnamefont{S.}~\bibnamefont{Krieg}},
  \bibinfo{author}{\bibfnamefont{J.}~\bibnamefont{Laeuchli}},
  \bibinfo{author}{\bibfnamefont{J.}~\bibnamefont{Negele}},
  \bibinfo{author}{\bibfnamefont{K.}~\bibnamefont{Orginos}},
  \bibinfo{author}{\bibfnamefont{A.}~\bibnamefont{Pochinsky}},
  \bibnamefont{and} \bibinfo{author}{\bibfnamefont{S.}~\bibnamefont{Syritsyn}},
  \bibinfo{journal}{Phys. Rev.} \textbf{\bibinfo{volume}{D92}},
  \bibinfo{pages}{031501(R)} (\bibinfo{year}{2015}), \eprint{1505.01803}.

\bibitem[{\citenamefont{Alexandrou et~al.}(2018)\citenamefont{Alexandrou,
  Constantinou, Hadjiyiannakou, Jansen, Kallidonis, Koutsou, and Vaquero
  Avil{\'e}s-Casco}}]{Alexandrou:2018zdf}
\bibinfo{author}{\bibfnamefont{C.}~\bibnamefont{Alexandrou}},
  \bibinfo{author}{\bibfnamefont{M.}~\bibnamefont{Constantinou}},
  \bibinfo{author}{\bibfnamefont{K.}~\bibnamefont{Hadjiyiannakou}},
  \bibinfo{author}{\bibfnamefont{K.}~\bibnamefont{Jansen}},
  \bibinfo{author}{\bibfnamefont{C.}~\bibnamefont{Kallidonis}},
  \bibinfo{author}{\bibfnamefont{G.}~\bibnamefont{Koutsou}}, \bibnamefont{and}
  \bibinfo{author}{\bibfnamefont{A.V.}~\bibnamefont{Avil{\'e}s-Casco}},
  \bibinfo{journal}{Phys. Rev.} \textbf{\bibinfo{volume}{D97}},
  \bibinfo{pages}{094504} (\bibinfo{year}{2018}), \eprint{1801.09581}.

\bibitem[{\citenamefont{Edwards and Joo}(2005)}]{Edwards:2004sx}
\bibinfo{author}{\bibfnamefont{R.~G.} \bibnamefont{Edwards}} \bibnamefont{and}
  \bibinfo{author}{\bibfnamefont{B.}~\bibnamefont{Joo}}
  (\bibinfo{collaboration}{SciDAC, LHPC, UKQCD}), \bibinfo{journal}{Nucl. Phys.
  Proc. Suppl.} \textbf{\bibinfo{volume}{140}}, \bibinfo{pages}{832}
  (\bibinfo{year}{2005}), \bibinfo{note}{[,832(2004)]},
  \eprint{hep-lat/0409003}.

\bibitem[{\citenamefont{Luscher and Schaefer}(2013)}]{Luscher:2012av}
\bibinfo{author}{\bibfnamefont{M.}~\bibnamefont{L\"uscher}} \bibnamefont{and}
  \bibinfo{author}{\bibfnamefont{S.}~\bibnamefont{Schaefer}},
  \bibinfo{journal}{Comput. Phys. Commun.} \textbf{\bibinfo{volume}{184}},
  \bibinfo{pages}{519} (\bibinfo{year}{2013}), \eprint{1206.2809}.

\bibitem[{\citenamefont{Djukanovic}(2016)}]{Djukanovic:2016spv}
\bibinfo{author}{\bibfnamefont{D.}~\bibnamefont{Djukanovic}}
  (\bibinfo{year}{2016}), \eprint{1603.01576}.

\end{thebibliography}
\appendix

\onecolumngrid
\section{Supplemental Material}
For convenience we attach the  supplemental material to the published Letter in the following sections.
\section{Finite-Volume Dependence}
\begin{figure}[h]
	\includegraphics{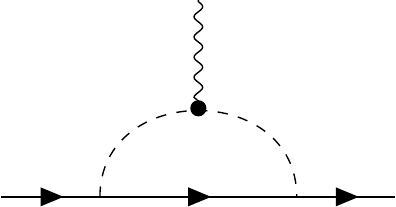}
	\caption{One-loop contribution to the strange magnetic moment.}\label{fig_diag}
\end{figure}

In this section we derive the finite-volume dependence of the strange magnetic
moment $\mu_s$ of the nucleon in HBChPT to order $\mathcal{O}(q^3)$. We will
show that the form of the finite-volume correction is the same as  in  the
SU(2) case for the isovector magnetic moment \cite{Beane:2004tw} after
substituting 
the kaon for the pion mass.  To this end we
analyze the relevant diagram in HBChPT \cite{Hemmert:1999mr}.
Only one diagram contributes to the magnetic moment at one loop to order $\mathcal{O}(q^3)$, see Fig.~\ref{fig_diag}.

The relevant meson-baryon Lagrangian is \cite{Bernard:1995dp}
\begin{align}
	\mathcal{L}&= D\left < \overline{B} S^\mu \{ u_\mu , B\} \right> + F\left< \overline{B} S^\mu [ u_\mu , B] \right>.
\end{align}

Expanding the Lagrangian in terms of the meson fields we obtain
\begin{align}
	\mathcal{L}&=\frac{1}{2F_\phi} D \left<\overline{B}_c \lambda_c S^\mu i \partial_\mu \phi_a \{\lambda_a,\lambda_b\} B_b\right>+  \frac{1}{2F_\phi}F \left<\overline{B}_c \lambda_c S^\mu i \partial_\mu \phi_a [\lambda_a,\lambda_b] B_b\right> + \dots \nonumber\\
	&=  2 D  d^{abc} \overline{B}_c  S^\mu i \partial_\mu \phi_a B_b+ 
	2 F i  f^{abc} \overline{B}_c  S^\mu i \partial_\mu \phi_a B_b +\dots
\end{align}
where we only show the terms necessary for the discussion of the finite-volume
effects. The $\lambda_i$ are the Gell-Mann matrices and the $d$ and $f$ are the
usual SU(3) structure functions.
This leads to the Feynman rule
\begin{align}
	\frac{2 ip\cdot S (D d^{abc} +  i F f^{abc})}{F_\phi},
\end{align}
for the meson-baryon interaction, where $p$ is the incoming momentum of the
meson with isospin index $a$, and $b,c$ are the isospin indices of the
incoming and outgoing baryon, respectively. The baryon propagator is given by
\begin{align}
	\frac{i}{v\cdot p}
\end{align}
The covariant derivative of the mesonic Lagrangian is defined as
\begin{align}
	D_\mu A &=  \partial_\mu A -i r_\mu A + i A l_\mu,\\
	r_\mu&= l_\mu=\lambda_8 v^{(8)}_\mu,
\end{align}
where for the magnetic moment only the octet current contributes
at the one-loop level. Again expanding the Lagrangian in terms of meson fields and only
keeping the relevant terms gives
\begin{align}
\mathcal{L}&= \frac{F^2}{4} \left< D_\mu U (D^\mu U )^\dagger\right>+\dots\nonumber\\
&= \frac{i}{2}\left<\partial_\mu \phi [\phi,\lambda^8]\right>  v^{(8)}+\dots\nonumber\\
&= -2 \partial_\mu \phi_a  \phi_b   f^{ab8}  v^{(8)} +\dots
\end{align}

The Feynman rule for the electromagnetic interaction of the meson reads
\begin{align}
	-2 f^{ab8}(p_\mu+p'_\mu)
\end{align}
where $p,a$ and $p',b$ are the momenta and isospin indices of the incoming and
outgoing meson, respectively. Since the structure functions $f^{ab8}$ only give
non-vanishing contributions for $a,b=4,5$ and $a,b=6,7$, only kaons contribute
to the loop diagram for the strange magnetic moment of the nucleon.
The matrix element of the current (in the Breit frame) is parametrized as
\begin{align}
	J_\mu&= \frac{1}{N_i N_f} \bar{u}(p') P^+ \Bigl [ 
	G_E v_\mu +\frac{1}{m} G_M [S_\mu,S_\nu] q^\nu 
\Bigr]P^+ u(p)
\end{align}
where
\begin{align}
	P^+&= \frac{1+\slashed{v}}{2} , 
	\quad\quad \quad \ \  q =  p' -p ,\quad\quad\ \ 
N_i = \sqrt{\frac{E_i+m}{2m}},\\
v^\mu&= \{1,0,0,0\},\quad 
p'^\mu= m v^\mu + r'^\mu,\quad
p^\mu= m v^\mu + r^\mu.
\end{align}
In the Breit frame the kinematic vectors read
\begin{align}
	r'^\mu&= \{E-m,\frac{\mathbf{q}}{2}\},\\
	r^\mu&= \{ E-m, -\frac{\mathbf{q}}{2}\},\\
	q^\mu&=r'^\mu-r^\mu= \{0,\mathbf{q}\},\\
	v^2 &=  1,\\
	v\cdot q &=  0.
\end{align}
Using the explicit representation of $S$
\begin{align}
	S_\mu&= \frac{i}{2} \gamma_5 \sigma^{\mu\nu} v^\nu,
\end{align}
we find that the part of a diagram proportional to $\gamma_\mu$ corresponds to the magnetic moment.

The one-loop  diagram of Fig.~\ref{fig_diag} reads
\begin{align}
	D&=   \frac{8m \Theta_{ba}}{F_\phi^2}\frac{1}{i} \int \frac{d^D k }{(2\pi)^D}\frac{k^\mu \slashed{k} }{(v \cdot k -\omega) ( k^2 -m_K^2) ((k+q)^2 -m_K^2)}+\dots ,
\end{align}
where we only display terms proportional to $\gamma_\mu$, i.e.  contributing to the magnetic moment. We have collected the isospin-dependent part in $\Theta$,
\begin{align}
	\Theta_{ba}&= -i (D d^{c a e} + i F f^{c a e}) f^{c d 
	8} ( D d^{d e b} + i  F f^{d e b}) ,
\end{align}
with $a,b$ the isospin index of the incoming, outgoing nucleon, respectively,
and $\omega=v\cdot p$.
We parametrize the tensor integral
\begin{align}
	\frac{1}{i}\int \frac{d^Dk}{(2\pi)^D} \frac{k^\mu k^\nu } {(v \cdot k -\omega) ( k^2 -m_K^2) ((k+q)^2 -m_K^2)} &=  g^{\mu\nu} c_1+ q^\mu q^\nu c_2 + (v^\mu q^\nu+ v^\nu q^\mu) c_3 + v^\mu v^\nu c_4.
\end{align}
For the subsequent discussion we only need $c_1$ which for the case $q^2=0$ and $v\cdot q=0$ reads\footnote{Note that $D$ here refers to space-time dimensions.}
\begin{align}
	c_1&= \frac{1}{D-1} \Bigl[ (m_K^2-\omega^2) K_0 + J_0 -\omega I_0\Bigr],
	\intertext{with}
	K_0&= \frac{1}{i}\int \frac{d^D k }{(2\pi)^D}\frac{1}{(v \cdot k -\omega) ( k^2 -m_K^2) ((k+q)^2 -m_K^2)},\\
	J_0&= \frac{1}{i}\int \frac{d^D k }{(2\pi)^D}\frac{1}{(v \cdot k -\omega) ( k^2 -m_K^2)},\\
	I_0&= \frac{1}{i}\int \frac{d^D k }{(2\pi)^D}\frac{1}{( k^2 -m_K^2) ((k+q)^2 -m_K^2)}.
\end{align}
The magnetic moment to one loop reads 
\begin{align}
	G^{(8),\text{loop}}_M(0,b,a)&= \frac{8 m \Theta_{ba}}{F_\phi^2} \frac{1}{D-1} \Bigl[ (m_K^2-\omega^2) K_0 + J_0 -\omega I_0\Bigr], \nonumber\\
	&= \frac{8 m \Theta_{ba}}{F_\phi^2} \frac{1}{D-1} \frac{\partial}{\partial m_K^2}  \underbrace{\Bigl[ (m_K^2-\omega^2) J_0 - \omega A_0\Bigr]}_{\Sigma} ,
		\intertext{with}
		A_0&= \frac{1}{i}\int \frac{d^D k }{(2\pi)^D}\frac{1}{( k^2 -m_K^2)}.
\end{align}
The isospin factor for the nucleon is
\begin{align}
	\Theta_{44}-i \Theta_{54}&= -\frac{5 D^2-6 D F+9 F^2}{4 \sqrt{3}}.
\end{align}
Inserting the explicit expression for the loop integrals, e.g. Appendix B in Ref. \cite{Bernard:1995dp}, we obtain
\begin{align}
	G^{(8),\text{loop}}_M(0)&= -\frac{m m_K}{8 \sqrt{3} \pi F_\phi^2}\Bigl(5 D^2-6 D F+9 F^2\Bigr),
\end{align}
which is the same result as in Ref. \cite{Hemmert:1999mr}.
Thus we have shown that the magnetic moment is proportional to the derivative of the self-energy with respect to $m_K^2$. Furthermore, we can rewrite $\Sigma$
\begin{align}
	\Bigl[ (m_K^2-\omega^2) J_0 - \omega A_0\Bigr] &= \frac{1}{i}\int \frac{d^D k}{(2\pi)^D} \frac{(m_K^2 - \omega^2) - \omega (v\cdot k -\omega) }{(v\cdot k -\omega) (k^2 -m_K^2)},\nonumber\\
	&=\frac{1}{i}  \int \frac{d^D k}{(2\pi)^D} \frac{k^2 - (v\cdot k)^2}{(v\cdot k -\omega) (k^2 -m_K^2)},\nonumber\\
	&= - \frac{1}{i} \int \frac{d^D k}{(2\pi)^D} \frac{\vec{k}^2}{(v\cdot k -\omega) (k^2 -m_K^2)}.
\end{align}
This expression coincides with the integral of Eq.~(8) from Ref.~\cite{Beane:2004tw} (up to an irrelevant factor), with the kaon mass
substituted for the pion mass. Thus the finite-volume corrections for the
strange magnetic moment of the nucleon are of the same form as in \cite{Beane:2004tw}, after substituting the kaon for the pion mass.
\clearpage

\section{Effective Mass}
\begin{figure}[h]
\includegraphics{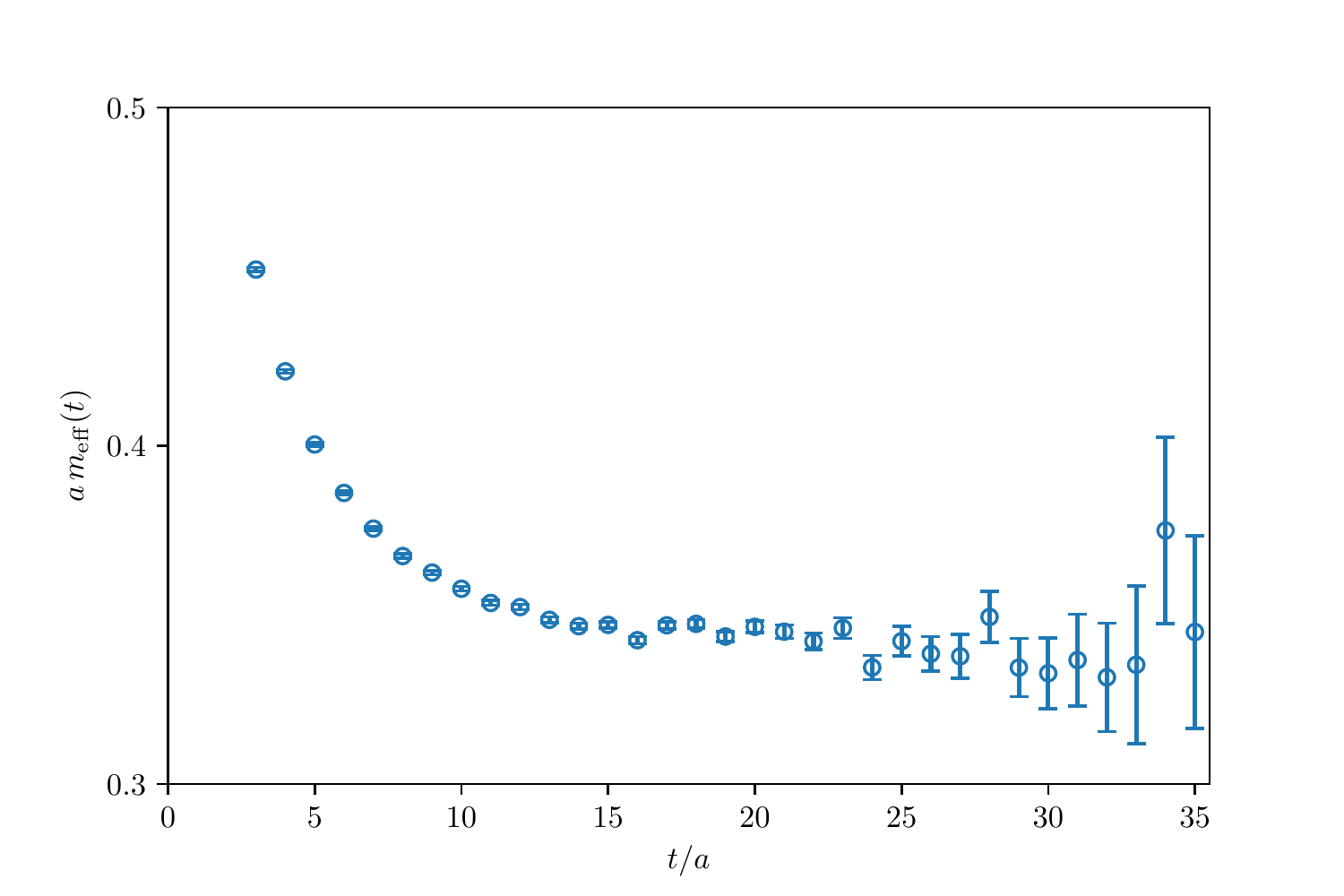}
\caption{Effective mass of the nucleon at zero momentum for ensemble N200.}\label{supp_fig_eff_mass}
\end{figure}

For convenience we show the effective mass of the nucleon for ensemble N200 at
zero momentum in Fig.~\ref{supp_fig_eff_mass} 

\section{Tables}
In this section we give the extracted form factors $G_{E/M}^s$ as well as the $z$-expansion fits for the final result quoted in the main text.

\begin{table}[h]
\begin{tabular}{|c|c|c|c|c|}
\hline
H105	&\multicolumn{2}{c|}{$G_E^s$}	&\multicolumn{2}{c|}{$G_M^s$}\\
\hline
$a_k$ &Summation Method	&Plateau Fit &Summation Method	&Plateau Fit\\
\hline
0	&-	&-	&-0.02047 (0.00437)	&-0.01233 (0.00599)\\
1	&0.06397 (0.01385)	&0.06823 (0.01405)	&0.12329 (0.06527)	&0.04532 (0.07789)\\
2	&0.02234 (0.14166)	&-0.02693 (0.14610)	&-0.00992 (0.28592)	&0.00175 (0.23002)\\
3	&0.00627 (0.15825)	&-0.00133 (0.16380)	&-0.00232 (0.28851)	&0.00026 (0.23533)\\
4	&0.00108 (0.15816)	&0.00019 (0.16458)	&-0.00038 (0.28453)	&0.00002 (0.23424)\\
5	&0.00015 (0.15918)	&0.00006 (0.16606)	&-0.00005 (0.28363)	&-0.00000 (0.23054)\\
\hline
$\chi^2$/dof	&0.44508	&1.64311	&1.46231	&0.48937\\
\hline
\end{tabular}
\caption{Fit of the $z$-expansion to the strange electromagnetic form factors on ensemble H105 with a transferred four-momentum cut of $Q^2<0.5\,\text{GeV}^2$.}
\end{table}

\begin{table}[h]
\begin{tabular}{|c|c|c|c|c|}
\hline
N401	&\multicolumn{2}{c|}{$G_E^s$}	&\multicolumn{2}{c|}{$G_M^s$}\\
\hline
$a_k$ &Summation Method	&Plateau Fit &Summation Method	&Plateau Fit\\
\hline
0	&-	&-	&-0.02529 (0.00425)	&-0.02095 (0.00435)\\
1	&0.09512 (0.01429)	&0.10383 (0.01443)	&0.13391 (0.09905)	&0.13431 (0.08529)\\
2	&-0.27623 (0.16770)	&-0.30477 (0.16933)	&0.13998 (0.62712)	&-0.04352 (0.48865)\\
3	&-0.02203 (0.25734)	&-0.03529 (0.26595)	&0.01803 (0.67666)	&-0.01013 (0.49372)\\
4	&-0.00082 (0.25992)	&-0.00313 (0.26699)	&0.00121 (0.68362)	&-0.00160 (0.49435)\\
5	&0.00007 (0.25824)	&-0.00025 (0.26746)	&-0.00001 (0.67480)	&-0.00021 (0.49920)\\
\hline
$\chi^2$/dof	&1.73239	&1.16637	&1.47959	&0.96105\\
\hline
\end{tabular}
\caption{Fit of the $z$-expansion to the strange electromagnetic form factors on ensemble N401 with a transferred four-momentum cut of $Q^2<0.5\,\text{GeV}^2$.}
\end{table}

\begin{table}[h]
\begin{tabular}{|c|c|c|c|c|}
\hline
N203	&\multicolumn{2}{c|}{$G_E^s$}	&\multicolumn{2}{c|}{$G_M^s$}\\
\hline
$a_k$ &Summation Method	&Plateau Fit &Summation Method	&Plateau Fit\\
\hline
0	&-	&-	&-0.01899 (0.00345)	&-0.01435 (0.00458)\\
1	&0.07188 (0.00981)	&0.06983 (0.01178)	&0.06979 (0.06544)	&0.02585 (0.06573)\\
2	&-0.24568 (0.09677)	&-0.22658 (0.11057)	&0.00194 (0.34785)	&-0.03506 (0.28291)\\
3	&-0.03558 (0.15444)	&-0.03115 (0.14710)	&-0.00407 (0.36546)	&-0.00927 (0.28600)\\
4	&-0.00422 (0.15360)	&-0.00345 (0.14939)	&-0.00127 (0.37107)	&-0.00165 (0.28557)\\
5	&-0.00049 (0.15529)	&-0.00037 (0.14689)	&-0.00025 (0.37221)	&-0.00025 (0.28763)\\
\hline
$\chi^2$/dof	&1.89473	&1.84721	&1.71021	&1.18630\\
\hline
\end{tabular}
\caption{Fit of the $z$-expansion to the strange electromagnetic form factors on ensemble N203 with a transferred four-momentum cut of $Q^2<0.5\,\text{GeV}^2$.}
\end{table}

\begin{table}[h]
\begin{tabular}{|c|c|c|c|c|}
\hline
N200	&\multicolumn{2}{c|}{$G_E^s$}	&\multicolumn{2}{c|}{$G_M^s$}\\
\hline
$a_k$ &Summation Method	&Plateau Fit &Summation Method	&Plateau Fit\\
\hline
0	&-	&-	&-0.02586 (0.00341)	&-0.02665 (0.00517)\\
1	&0.08026 (0.01134)	&0.06574 (0.01291)	&0.20748 (0.06348)	&0.23425 (0.06965)\\
2	&-0.34096 (0.11673)	&-0.19624 (0.13298)	&-0.16836 (0.34266)	&-0.04257 (0.23831)\\
3	&-0.06001 (0.15726)	&-0.03532 (0.16129)	&-0.04070 (0.35692)	&-0.01019 (0.24036)\\
4	&-0.00831 (0.15976)	&-0.00498 (0.16238)	&-0.00670 (0.35371)	&-0.00166 (0.23809)\\
5	&-0.00106 (0.15881)	&-0.00064 (0.16369)	&-0.00094 (0.35217)	&-0.00023 (0.23128)\\
\hline
$\chi^2$/dof	&1.69598	&0.96513	&0.96210	&1.88794\\
\hline
\end{tabular}
\caption{Fit of the $z$-expansion to the strange electromagnetic form factors on ensemble N200 with a transferred four-momentum cut of $Q^2<0.5\,\text{GeV}^2$.}
\end{table}

\begin{table}[h]
\begin{tabular}{|c|c|c|c|c|}
\hline
D200	&\multicolumn{2}{c|}{$G_E^s$}	&\multicolumn{2}{c|}{$G_M^s$}\\
\hline
$a_k$ &Summation Method	&Plateau Fit &Summation Method	&Plateau Fit\\
\hline
0	&-	&-	&-0.01544 (0.00470)	&-0.01214 (0.00862)\\
1	&0.06857 (0.02031)	&0.06464 (0.02218)	&0.10160 (0.07439)	&0.09163 (0.13288)\\
2	&-0.01483 (0.22268)	&-0.14348 (0.24059)	&-0.06996 (0.37373)	&-0.00272 (0.63158)\\
3	&0.00768 (0.31098)	&-0.01647 (0.28613)	&-0.01276 (0.36976)	&0.00061 (0.64526)\\
4	&0.00191 (0.30663)	&-0.00143 (0.28985)	&-0.00165 (0.36929)	&0.00019 (0.64590)\\
5	&0.00030 (0.30945)	&-0.00010 (0.28545)	&-0.00019 (0.37093)	&0.00003 (0.65096)\\
\hline
$\chi^2$/dof	&1.05330	&1.14854	&1.77733	&0.59909\\
\hline
\end{tabular}
\caption{Fit of the $z$-expansion to the strange electromagnetic form factors on ensemble D200 with a transferred four-momentum cut of $Q^2<0.5\,\text{GeV}^2$.}
\end{table}

\begin{table}[h]
\begin{tabular}{|c|c|c|c|c|}
\hline
N302	&\multicolumn{2}{c|}{$G_E^s$}	&\multicolumn{2}{c|}{$G_M^s$}\\
\hline
$a_k$ &Summation Method	&Plateau Fit &Summation Method	&Plateau Fit\\
\hline
0	&-	&-	&-0.01088 (0.00320)	&-0.00910 (0.00515)\\
1	&0.06065 (0.00886)	&0.05927 (0.01035)	&0.01563 (0.04561)	&0.00821 (0.06148)\\
2	&-0.13785 (0.07965)	&-0.18502 (0.09230)	&-0.00556 (0.19990)	&-0.00875 (0.19448)\\
3	&-0.02385 (0.09626)	&-0.03279 (0.10684)	&-0.00165 (0.19593)	&-0.00221 (0.19935)\\
4	&-0.00316 (0.09728)	&-0.00449 (0.10823)	&-0.00033 (0.19528)	&-0.00038 (0.20066)\\
5	&-0.00038 (0.09759)	&-0.00056 (0.11077)	&-0.00005 (0.19247)	&-0.00006 (0.19969)\\
\hline
$\chi^2$/dof	&2.72651	&1.64723	&1.55374	&2.20057\\
\hline
\end{tabular}
\caption{Fit of the $z$-expansion to the strange electromagnetic form factors on ensemble N302 with a transferred four-momentum cut of $Q^2<0.5\,\text{GeV}^2$.}
\end{table}

\begin{table}[h]
\begin{tabular}{|c|c|c|c|c|}
\hline
H105	&\multicolumn{2}{c|}{$G_E^s$}	&\multicolumn{2}{c|}{$G_M^s$}\\
\hline
$Q^2\,[\text{GeV}^2]$ &Summation Method	&Plateau Fit &Summation Method	&Plateau Fit\\
\hline
0.00000	&0.00034 (0.00107)	&0.00135 (0.00106)	& - 	& - \\
0.14300	&0.00327 (0.00155)	&0.00195 (0.00233)	&-0.02275 (0.00903)	&-0.00662 (0.01681)\\
0.14974	&0.00304 (0.00156)	&0.00585 (0.00197)	&-0.01549 (0.00658)	&-0.00950 (0.01191)\\
0.19268	&0.00346 (0.00080)	&0.00378 (0.00077)	&-0.01174 (0.00267)	&-0.00795 (0.00374)\\
0.19397	&0.00350 (0.00067)	&0.00383 (0.00069)	&-0.01517 (0.00226)	&-0.01139 (0.00306)\\
0.19487	&0.00408 (0.00093)	&0.00557 (0.00113)	&-0.00947 (0.00324)	&-0.01279 (0.00545)\\
0.30464	&0.00399 (0.00154)	&0.00532 (0.00211)	&-0.02289 (0.00514)	&-0.02235 (0.00939)\\
0.31545	&0.00406 (0.00191)	&0.00576 (0.00275)	&-0.01645 (0.00510)	&-0.02176 (0.01025)\\
0.37069	&0.00544 (0.00065)	&0.00423 (0.00081)	&-0.00832 (0.00166)	&-0.00677 (0.00251)\\
0.37505	&0.00537 (0.00093)	&0.00592 (0.00124)	&-0.01291 (0.00269)	&-0.01148 (0.00406)\\
0.37833	&0.00529 (0.00115)	&0.00605 (0.00188)	&-0.01197 (0.00295)	&-0.00765 (0.00611)\\
0.40252	&0.00644 (0.00064)	&0.00666 (0.00076)	&-0.00987 (0.00148)	&-0.00880 (0.00213)\\
0.45865	&0.00495 (0.00340)	&0.00744 (0.00375)	&-0.01160 (0.00850)	&-0.01750 (0.01360)\\
0.53690	&0.00486 (0.00170)	&0.00337 (0.00243)	&-0.00795 (0.00313)	&-0.01109 (0.00592)\\
0.55227	&0.00488 (0.00157)	&0.00524 (0.00189)	&-0.00937 (0.00308)	&-0.00146 (0.00521)\\
0.59650	&0.00496 (0.00083)	&0.00458 (0.00094)	&-0.00896 (0.00172)	&-0.00855 (0.00219)\\
0.69338	&0.00541 (0.00262)	&0.00305 (0.00435)	&-0.00433 (0.00486)	&0.01567 (0.00764)\\
0.70727	&0.00323 (0.00238)	&-0.00021 (0.00309)	&-0.00360 (0.00459)	&-0.00018 (0.00773)\\
0.71798	&0.00484 (0.00308)	&0.00288 (0.00441)	&-0.00341 (0.00528)	&0.00410 (0.01151)\\
0.80515	&0.00473 (0.00126)	&0.00472 (0.00158)	&-0.00471 (0.00199)	&-0.00449 (0.00308)\\
0.84184	&0.00525 (0.00178)	&0.00582 (0.00258)	&0.00120 (0.00339)	&-0.00045 (0.00599)\\
0.86127	&0.00340 (0.00207)	&0.00559 (0.00252)	&-0.00265 (0.00363)	&-0.00752 (0.00603)\\
0.94815	&0.00594 (0.00132)	&0.00857 (0.00204)	&-0.00082 (0.00227)	&-0.00067 (0.00427)\\
0.95489	&0.00439 (0.00117)	&0.00710 (0.00166)	&-0.00245 (0.00190)	&-0.00146 (0.00349)\\
0.98315	&0.00046 (0.00370)	&0.00405 (0.00415)	&-0.00748 (0.00583)	&-0.00590 (0.00838)\\
0.99902	&0.00440 (0.00073)	&0.00557 (0.00086)	&-0.00336 (0.00108)	&-0.00403 (0.00162)\\
1.00001	&0.00411 (0.00093)	&0.00390 (0.00127)	&-0.00285 (0.00143)	&0.00055 (0.00260)\\
1.10979	&0.00439 (0.00182)	&0.00403 (0.00232)	&-0.00636 (0.00258)	&-0.00900 (0.00425)\\
1.12050	&0.00200 (0.00270)	&0.00614 (0.00378)	&-0.01003 (0.00381)	&-0.00675 (0.00683)\\
1.18020	&0.00310 (0.00135)	&0.00278 (0.00207)	&-0.00531 (0.00177)	&-0.00186 (0.00353)\\
1.18347	&0.00606 (0.00169)	&0.00209 (0.00272)	&-0.00069 (0.00232)	&0.00083 (0.00451)\\
1.20767	&0.00472 (0.00096)	&0.00397 (0.00134)	&-0.00422 (0.00133)	&-0.00342 (0.00189)\\

\hline
\end{tabular}
\caption{Results for the strange electromagnetic form factors from the summation method and plateau fit at $1\,\text{fm}$ on ensemble H105.}
\end{table}

\begin{table}
\begin{tabular}{|c|c|c|c|c|}
\hline
N401	&\multicolumn{2}{c|}{$G_E^s$}	&\multicolumn{2}{c|}{$G_M^s$}\\
\hline
$Q^2\,[\text{GeV}^2]$ &Summation Method	&Plateau Fit &Summation Method	&Plateau Fit\\
\hline
0.00000	&0.00116 (0.00115)	&0.00176 (0.00112)	& - 	& - \\
0.09297	&0.00256 (0.00135)	&0.00255 (0.00128)	&0.00571 (0.00949)	&-0.01468 (0.01318)\\
0.09455	&0.00238 (0.00126)	&0.00284 (0.00123)	&-0.02148 (0.00782)	&-0.01459 (0.00972)\\
0.11160	&0.00353 (0.00072)	&0.00371 (0.00072)	&-0.01811 (0.00394)	&-0.02243 (0.00448)\\
0.11190	&0.00404 (0.00066)	&0.00366 (0.00063)	&-0.02018 (0.00337)	&-0.01543 (0.00396)\\
0.11209	&0.00399 (0.00082)	&0.00418 (0.00087)	&-0.01899 (0.00413)	&-0.01054 (0.00517)\\
0.19208	&0.00273 (0.00118)	&0.00442 (0.00109)	&-0.01577 (0.00571)	&-0.01013 (0.00684)\\
0.19485	&0.00321 (0.00122)	&0.00517 (0.00125)	&-0.01864 (0.00481)	&-0.01690 (0.00558)\\
0.21804	&0.00444 (0.00061)	&0.00543 (0.00059)	&-0.02050 (0.00219)	&-0.01470 (0.00280)\\
0.21904	&0.00349 (0.00081)	&0.00563 (0.00080)	&-0.01587 (0.00311)	&-0.01153 (0.00379)\\
0.21983	&0.00278 (0.00085)	&0.00412 (0.00091)	&-0.01489 (0.00374)	&-0.00720 (0.00490)\\
0.22895	&0.00407 (0.00061)	&0.00495 (0.00059)	&-0.01420 (0.00214)	&-0.01249 (0.00234)\\
0.28782	&0.00237 (0.00200)	&0.00585 (0.00184)	&-0.02224 (0.00656)	&-0.01464 (0.00727)\\
0.31993	&0.00327 (0.00138)	&0.00644 (0.00127)	&-0.01616 (0.00366)	&-0.01296 (0.00478)\\
0.32360	&0.00350 (0.00128)	&0.00694 (0.00118)	&-0.01782 (0.00372)	&-0.00876 (0.00433)\\
0.34084	&0.00539 (0.00084)	&0.00640 (0.00079)	&-0.01255 (0.00228)	&-0.00853 (0.00258)\\
0.41775	&0.00395 (0.00200)	&0.00430 (0.00184)	&-0.00471 (0.00469)	&-0.01755 (0.00620)\\
0.42102	&0.00312 (0.00200)	&0.00509 (0.00184)	&-0.00454 (0.00446)	&-0.00760 (0.00546)\\
0.42380	&0.00313 (0.00218)	&0.00520 (0.00212)	&-0.00826 (0.00454)	&-0.01159 (0.00641)\\
0.45789	&0.00629 (0.00118)	&0.00684 (0.00114)	&-0.00927 (0.00283)	&-0.00828 (0.00338)\\
0.51211	&0.00633 (0.00122)	&0.00475 (0.00110)	&-0.00534 (0.00317)	&-0.00501 (0.00336)\\
0.51687	&0.00663 (0.00129)	&0.00522 (0.00116)	&-0.00912 (0.00287)	&-0.00603 (0.00326)\\
0.55086	&0.00599 (0.00102)	&0.00509 (0.00096)	&-0.00268 (0.00249)	&-0.00478 (0.00300)\\
0.55255	&0.00631 (0.00087)	&0.00620 (0.00086)	&-0.00577 (0.00193)	&-0.00728 (0.00233)\\
0.56979	&0.00614 (0.00059)	&0.00550 (0.00062)	&-0.00607 (0.00147)	&-0.00421 (0.00169)\\
0.56999	&0.00677 (0.00074)	&0.00662 (0.00076)	&-0.00809 (0.00164)	&-0.00626 (0.00201)\\
0.60319	&0.00327 (0.00176)	&0.00315 (0.00167)	&-0.00668 (0.00370)	&-0.00463 (0.00425)\\
0.64997	&0.00378 (0.00122)	&0.00377 (0.00117)	&-0.00705 (0.00261)	&-0.00771 (0.00304)\\
0.65275	&0.00278 (0.00154)	&0.00343 (0.00152)	&-0.01197 (0.00331)	&-0.00859 (0.00380)\\
0.67693	&0.00438 (0.00102)	&0.00455 (0.00098)	&-0.00664 (0.00194)	&-0.00372 (0.00247)\\
0.67772	&0.00494 (0.00122)	&0.00396 (0.00131)	&-0.00835 (0.00248)	&-0.00954 (0.00295)\\
0.68694	&0.00512 (0.00083)	&0.00558 (0.00082)	&-0.00497 (0.00157)	&-0.00342 (0.00184)\\
\hline
\end{tabular}
\caption{Results for the strange electromagnetic form factors from the summation method and plateau fit at $1\,\text{fm}$ on ensemble N401.}
\end{table}

\begin{table}
\begin{tabular}{|c|c|c|c|c|}
\hline
N203	&\multicolumn{2}{c|}{$G_E^s$}	&\multicolumn{2}{c|}{$G_M^s$}\\
\hline
$Q^2\,[\text{GeV}^2]$ &Summation Method	&Plateau Fit &Summation Method	&Plateau Fit\\
\hline
0.00000	&-0.00003 (0.00090)	&-0.00153 (0.00112)	& - 	& - \\
0.12555	&0.00308 (0.00102)	&0.00534 (0.00123)	&-0.00476 (0.00624)	&-0.01342 (0.01149)\\
0.12876	&0.00289 (0.00096)	&0.00532 (0.00121)	&-0.00834 (0.00493)	&-0.00039 (0.00845)\\
0.15658	&0.00455 (0.00052)	&0.00515 (0.00069)	&-0.01772 (0.00258)	&-0.01591 (0.00487)\\
0.15718	&0.00425 (0.00047)	&0.00481 (0.00064)	&-0.01598 (0.00204)	&-0.01474 (0.00295)\\
0.15758	&0.00409 (0.00062)	&0.00502 (0.00080)	&-0.01677 (0.00288)	&-0.01592 (0.00403)\\
0.26184	&0.00222 (0.00088)	&0.00105 (0.00118)	&-0.00891 (0.00390)	&-0.00484 (0.00568)\\
0.26716	&0.00169 (0.00094)	&0.00067 (0.00140)	&-0.01402 (0.00331)	&-0.01198 (0.00512)\\
0.30442	&0.00357 (0.00041)	&0.00337 (0.00062)	&-0.01435 (0.00139)	&-0.01170 (0.00233)\\
0.30633	&0.00324 (0.00059)	&0.00231 (0.00079)	&-0.01329 (0.00197)	&-0.01114 (0.00295)\\
0.30794	&0.00329 (0.00063)	&0.00271 (0.00089)	&-0.01292 (0.00233)	&-0.01139 (0.00356)\\
0.32310	&0.00390 (0.00043)	&0.00321 (0.00057)	&-0.01245 (0.00121)	&-0.01160 (0.00188)\\
0.39291	&0.00613 (0.00172)	&0.00528 (0.00219)	&-0.00256 (0.00419)	&0.00367 (0.00728)\\
0.44463	&0.00597 (0.00102)	&0.00611 (0.00127)	&-0.01222 (0.00238)	&-0.01587 (0.00502)\\
0.45196	&0.00499 (0.00103)	&0.00509 (0.00128)	&-0.00788 (0.00224)	&-0.00743 (0.00363)\\
0.48029	&0.00502 (0.00053)	&0.00530 (0.00070)	&-0.01001 (0.00134)	&-0.01118 (0.00212)\\
0.57851	&0.00171 (0.00141)	&0.00175 (0.00165)	&-0.01045 (0.00336)	&-0.01087 (0.00506)\\
0.58494	&0.00150 (0.00137)	&0.00270 (0.00165)	&-0.00732 (0.00319)	&-0.00271 (0.00468)\\
0.59036	&0.00110 (0.00159)	&0.00381 (0.00209)	&-0.00874 (0.00341)	&-0.00602 (0.00521)\\
0.64631	&0.00410 (0.00077)	&0.00378 (0.00100)	&-0.00662 (0.00169)	&-0.00473 (0.00264)\\
0.70667	&0.00527 (0.00093)	&0.00570 (0.00117)	&-0.00508 (0.00198)	&-0.00962 (0.00344)\\
0.71601	&0.00516 (0.00098)	&0.00564 (0.00138)	&-0.00762 (0.00214)	&-0.01061 (0.00351)\\
0.77185	&0.00552 (0.00073)	&0.00506 (0.00099)	&-0.00312 (0.00159)	&-0.00774 (0.00256)\\
0.77507	&0.00540 (0.00063)	&0.00481 (0.00094)	&-0.00526 (0.00130)	&-0.00642 (0.00214)\\
0.80349	&0.00538 (0.00041)	&0.00513 (0.00058)	&-0.00574 (0.00086)	&-0.00660 (0.00139)\\
0.80389	&0.00528 (0.00052)	&0.00488 (0.00076)	&-0.00573 (0.00102)	&-0.00721 (0.00167)\\
0.82990	&0.00616 (0.00157)	&0.00483 (0.00182)	&-0.00194 (0.00270)	&-0.00748 (0.00448)\\
0.90814	&0.00444 (0.00092)	&0.00313 (0.00121)	&-0.00523 (0.00175)	&-0.00332 (0.00258)\\
0.91347	&0.00509 (0.00127)	&0.00514 (0.00177)	&-0.00354 (0.00226)	&0.00169 (0.00355)\\
0.95264	&0.00493 (0.00076)	&0.00517 (0.00097)	&-0.00426 (0.00117)	&-0.00516 (0.00192)\\
0.95424	&0.00397 (0.00083)	&0.00330 (0.00116)	&-0.00509 (0.00151)	&-0.00626 (0.00235)\\
0.96941	&0.00425 (0.00055)	&0.00417 (0.00072)	&-0.00515 (0.00090)	&-0.00531 (0.00147)\\
\hline
\end{tabular}
\caption{Results for the strange electromagnetic form factors from the summation method and plateau fit at $1\,\text{fm}$ on ensemble N203.}
\end{table}

\begin{table}
\begin{tabular}{|c|c|c|c|c|}
\hline
N200	&\multicolumn{2}{c|}{$G_E^s$}	&\multicolumn{2}{c|}{$G_M^s$}\\
\hline
$Q^2\,[\text{GeV}^2]$ &Summation Method	&Plateau Fit &Summation Method	&Plateau Fit\\
\hline
0.00000	&-0.00093 (0.00106)	&-0.00055 (0.00139)	& - 	& - \\
0.12293	&0.00313 (0.00116)	&0.00204 (0.00161)	&-0.01729 (0.00681)	&-0.02589 (0.01542)\\
0.12665	&0.00380 (0.00114)	&0.00154 (0.00163)	&-0.01656 (0.00600)	&0.00015 (0.01105)\\
0.15618	&0.00331 (0.00057)	&0.00255 (0.00079)	&-0.02071 (0.00244)	&-0.02093 (0.00519)\\
0.15678	&0.00343 (0.00054)	&0.00320 (0.00073)	&-0.01831 (0.00221)	&-0.02165 (0.00364)\\
0.15728	&0.00371 (0.00068)	&0.00297 (0.00097)	&-0.02179 (0.00303)	&-0.02532 (0.00509)\\
0.25772	&0.00447 (0.00103)	&0.00183 (0.00147)	&-0.01230 (0.00460)	&-0.01697 (0.00803)\\
0.26375	&0.00407 (0.00110)	&0.00205 (0.00175)	&-0.01123 (0.00402)	&-0.01699 (0.00742)\\
0.30282	&0.00451 (0.00047)	&0.00403 (0.00063)	&-0.00973 (0.00144)	&-0.00647 (0.00297)\\
0.30513	&0.00467 (0.00066)	&0.00365 (0.00092)	&-0.00858 (0.00259)	&-0.00497 (0.00401)\\
0.30693	&0.00483 (0.00073)	&0.00300 (0.00122)	&-0.01039 (0.00279)	&-0.01177 (0.00512)\\
0.32310	&0.00462 (0.00047)	&0.00445 (0.00063)	&-0.00893 (0.00149)	&-0.00491 (0.00246)\\
0.38688	&0.00871 (0.00207)	&0.00502 (0.00265)	&-0.00277 (0.00558)	&0.00434 (0.00987)\\
0.44152	&0.00677 (0.00127)	&0.00548 (0.00154)	&-0.00705 (0.00280)	&-0.01500 (0.00559)\\
0.44985	&0.00721 (0.00120)	&0.00418 (0.00160)	&-0.00347 (0.00281)	&0.00084 (0.00495)\\
0.47998	&0.00410 (0.00064)	&0.00359 (0.00090)	&-0.00628 (0.00161)	&-0.00191 (0.00270)\\
0.57339	&0.00402 (0.00175)	&0.00546 (0.00223)	&-0.00395 (0.00415)	&0.00219 (0.00841)\\
0.58082	&0.00504 (0.00167)	&0.00810 (0.00215)	&-0.00824 (0.00379)	&-0.00954 (0.00637)\\
0.58695	&0.00590 (0.00191)	&0.00857 (0.00279)	&-0.00484 (0.00420)	&-0.00907 (0.00725)\\
0.64631	&0.00444 (0.00090)	&0.00471 (0.00120)	&-0.00543 (0.00189)	&-0.00710 (0.00335)\\
0.69934	&0.00560 (0.00123)	&0.00744 (0.00153)	&-0.00483 (0.00234)	&-0.00775 (0.00467)\\
0.71008	&0.00702 (0.00127)	&0.00810 (0.00172)	&-0.00672 (0.00268)	&-0.00389 (0.00479)\\
0.76924	&0.00663 (0.00093)	&0.00643 (0.00129)	&-0.00345 (0.00192)	&-0.00493 (0.00360)\\
0.77296	&0.00686 (0.00081)	&0.00734 (0.00116)	&-0.00603 (0.00164)	&-0.00675 (0.00297)\\
0.80309	&0.00538 (0.00051)	&0.00581 (0.00074)	&-0.00520 (0.00100)	&-0.00521 (0.00187)\\
0.80359	&0.00580 (0.00065)	&0.00610 (0.00098)	&-0.00561 (0.00123)	&-0.00451 (0.00226)\\
0.82016	&0.00558 (0.00183)	&0.00182 (0.00251)	&-0.00303 (0.00299)	&-0.00551 (0.00552)\\
0.90403	&0.00412 (0.00110)	&0.00443 (0.00154)	&-0.00377 (0.00190)	&-0.00580 (0.00341)\\
0.91005	&0.00503 (0.00154)	&0.00296 (0.00233)	&-0.00446 (0.00257)	&-0.00640 (0.00468)\\
0.95133	&0.00393 (0.00086)	&0.00253 (0.00126)	&-0.00525 (0.00139)	&-0.00716 (0.00247)\\
0.95324	&0.00275 (0.00106)	&0.00381 (0.00157)	&-0.00352 (0.00180)	&-0.01072 (0.00386)\\
0.96941	&0.00418 (0.00063)	&0.00349 (0.00094)	&-0.00456 (0.00108)	&-0.00367 (0.00185)\\
\hline
\end{tabular}
\caption{Results for the strange electromagnetic form factors from the summation method and plateau fit at $1\,\text{fm}$ on ensemble N200.}
\end{table}

\begin{table}
\begin{tabular}{|c|c|c|c|c|}
\hline
D200	&\multicolumn{2}{c|}{$G_E^s$}	&\multicolumn{2}{c|}{$G_M^s$}\\
\hline
$Q^2\,[\text{GeV}^2]$ &Summation Method	&Plateau Fit &Summation Method	&Plateau Fit\\
\hline
0.00000	&0.00086 (0.00245)	&0.00279 (0.00314)	& - 	& - \\
0.07543	&-0.00185 (0.00223)	&0.00058 (0.00316)	&-0.02945 (0.01863)	&-0.02786 (0.03715)\\
0.07653	&-0.00117 (0.00205)	&0.00005 (0.00278)	&-0.01659 (0.01504)	&-0.01260 (0.02745)\\
0.08889	&0.00109 (0.00132)	&0.00313 (0.00165)	&-0.04089 (0.00895)	&-0.01793 (0.01775)\\
0.08899	&0.00204 (0.00113)	&0.00281 (0.00140)	&-0.01097 (0.00774)	&-0.01581 (0.01209)\\
0.08919	&0.00157 (0.00137)	&0.00168 (0.00191)	&-0.00955 (0.00956)	&-0.01169 (0.01477)\\
0.15527	&0.00327 (0.00198)	&0.00306 (0.00254)	&-0.03115 (0.01122)	&-0.02594 (0.01966)\\
0.15708	&0.00298 (0.00199)	&0.00419 (0.00258)	&-0.00891 (0.00922)	&-0.00692 (0.01470)\\
0.17406	&0.00319 (0.00110)	&0.00288 (0.00132)	&-0.00706 (0.00460)	&0.00191 (0.00896)\\
0.17466	&0.00421 (0.00143)	&0.00558 (0.00179)	&-0.00402 (0.00758)	&0.00724 (0.01104)\\
0.17516	&0.00381 (0.00156)	&0.00458 (0.00206)	&-0.02756 (0.00786)	&-0.02037 (0.01349)\\
0.18179	&0.00421 (0.00111)	&0.00455 (0.00129)	&-0.01730 (0.00493)	&-0.00671 (0.00716)\\
0.23251	&-0.00073 (0.00312)	&0.00008 (0.00429)	&-0.00278 (0.01129)	&-0.00205 (0.01903)\\
0.25591	&0.00279 (0.00238)	&0.00499 (0.00335)	&-0.00777 (0.00821)	&-0.02576 (0.01537)\\
0.25832	&0.00143 (0.00205)	&-0.00023 (0.00281)	&0.00007 (0.00698)	&-0.00774 (0.01103)\\
0.27078	&0.00365 (0.00134)	&0.00386 (0.00203)	&-0.00409 (0.00497)	&-0.01025 (0.00781)\\
0.33485	&-0.00071 (0.00318)	&0.00230 (0.00447)	&-0.00616 (0.00991)	&0.00424 (0.01819)\\
0.33696	&-0.00027 (0.00308)	&0.00356 (0.00401)	&-0.00806 (0.00987)	&0.00327 (0.01494)\\
0.33877	&-0.00012 (0.00322)	&0.00295 (0.00441)	&-0.00677 (0.00981)	&-0.00817 (0.01694)\\
0.36358	&0.00131 (0.00199)	&0.00179 (0.00264)	&-0.00295 (0.00648)	&-0.00256 (0.00979)\\
0.41129	&0.00664 (0.00220)	&0.00348 (0.00334)	&-0.01137 (0.00504)	&0.00449 (0.01013)\\
0.41430	&0.00626 (0.00208)	&0.00286 (0.00291)	&-0.00628 (0.00610)	&0.00222 (0.01017)\\
0.43901	&0.00609 (0.00176)	&0.00149 (0.00250)	&0.00627 (0.00490)	&-0.00887 (0.00899)\\
0.44001	&0.00696 (0.00142)	&0.00384 (0.00199)	&-0.00716 (0.00403)	&0.00015 (0.00635)\\
0.45257	&0.00690 (0.00110)	&0.00290 (0.00159)	&-0.00840 (0.00306)	&-0.00460 (0.00514)\\
0.45267	&0.00664 (0.00132)	&0.00448 (0.00199)	&-0.01535 (0.00388)	&-0.00104 (0.00621)\\
0.48521	&0.00882 (0.00274)	&0.01279 (0.00343)	&0.00182 (0.00625)	&0.00848 (0.01126)\\
0.51875	&0.00518 (0.00197)	&0.01002 (0.00237)	&-0.00284 (0.00512)	&0.00437 (0.00758)\\
0.52056	&0.00720 (0.00208)	&0.00733 (0.00290)	&0.00961 (0.00570)	&0.02815 (0.00982)\\
0.53814	&0.00770 (0.00164)	&0.01075 (0.00218)	&-0.00167 (0.00408)	&0.01878 (0.00710)\\
0.53874	&0.00379 (0.00197)	&0.01119 (0.00243)	&-0.00565 (0.00533)	&-0.00720 (0.00865)\\
0.54527	&0.00674 (0.00136)	&0.00909 (0.00167)	&-0.00219 (0.00337)	&0.00405 (0.00543)\\
\hline
\end{tabular}
\caption{Results for the strange electromagnetic form factors from the summation method and plateau fit at $1\,\text{fm}$ on ensemble D200.}
\end{table}

\begin{table}
\begin{tabular}{|c|c|c|c|c|}
\hline
N302	&\multicolumn{2}{c|}{$G_E^s$}	&\multicolumn{2}{c|}{$G_M^s$}\\
\hline
$Q^2\,[\text{GeV}^2]$ &Summation Method	&Plateau Fit &Summation Method	&Plateau Fit\\
\hline
0.00000	&0.00030 (0.00088)	&0.00098 (0.00099)	& - 	& - \\
0.18358	&0.00316 (0.00099)	&0.00582 (0.00153)	&-0.01073 (0.00529)	&-0.01557 (0.01293)\\
0.19423	&0.00389 (0.00096)	&0.00541 (0.00149)	&-0.00649 (0.00469)	&-0.01593 (0.00961)\\
0.25587	&0.00492 (0.00048)	&0.00460 (0.00060)	&-0.01281 (0.00146)	&-0.01527 (0.00273)\\
0.25800	&0.00451 (0.00044)	&0.00433 (0.00051)	&-0.00725 (0.00148)	&-0.00468 (0.00223)\\
0.25955	&0.00407 (0.00057)	&0.00425 (0.00081)	&-0.00570 (0.00209)	&-0.00661 (0.00381)\\
0.39513	&0.00320 (0.00102)	&0.00322 (0.00141)	&-0.00884 (0.00350)	&0.00210 (0.00711)\\
0.41206	&0.00453 (0.00130)	&0.00395 (0.00192)	&-0.00889 (0.00346)	&-0.00603 (0.00786)\\
0.49006	&0.00439 (0.00041)	&0.00363 (0.00052)	&-0.00899 (0.00084)	&-0.00878 (0.00167)\\
0.49732	&0.00484 (0.00058)	&0.00400 (0.00078)	&-0.00973 (0.00155)	&-0.00706 (0.00284)\\
0.50264	&0.00386 (0.00071)	&0.00247 (0.00107)	&-0.00983 (0.00176)	&-0.00365 (0.00384)\\
0.53787	&0.00444 (0.00041)	&0.00428 (0.00049)	&-0.00747 (0.00090)	&-0.00889 (0.00143)\\
0.59593	&0.00675 (0.00216)	&0.00583 (0.00305)	&0.00616 (0.00541)	&-0.00440 (0.01034)\\
0.70713	&0.00613 (0.00097)	&0.00583 (0.00125)	&-0.00791 (0.00189)	&-0.01123 (0.00383)\\
0.73200	&0.00547 (0.00096)	&0.00735 (0.00146)	&-0.00064 (0.00195)	&-0.00920 (0.00419)\\
0.79587	&0.00423 (0.00048)	&0.00471 (0.00065)	&-0.00583 (0.00102)	&-0.00616 (0.00164)\\
0.91055	&0.00491 (0.00159)	&0.00083 (0.00207)	&-0.00699 (0.00272)	&-0.00274 (0.00540)\\
0.93290	&0.00550 (0.00147)	&0.00451 (0.00199)	&-0.00165 (0.00271)	&0.00121 (0.00500)\\
0.94984	&0.00551 (0.00198)	&0.00179 (0.00320)	&-0.00200 (0.00323)	&0.00658 (0.00784)\\
1.07564	&0.00449 (0.00072)	&0.00409 (0.00094)	&-0.00340 (0.00118)	&-0.00387 (0.00201)\\
1.10255	&0.00445 (0.00119)	&0.00464 (0.00151)	&-0.00378 (0.00184)	&-0.00328 (0.00371)\\
1.13381	&0.00455 (0.00138)	&0.00466 (0.00202)	&-0.00100 (0.00243)	&0.00017 (0.00494)\\
1.25922	&0.00347 (0.00081)	&0.00439 (0.00119)	&-0.00292 (0.00142)	&0.00069 (0.00290)\\
1.26987	&0.00344 (0.00072)	&0.00409 (0.00118)	&-0.00077 (0.00117)	&0.00009 (0.00262)\\
1.28477	&0.00068 (0.00222)	&0.00171 (0.00280)	&0.00267 (0.00330)	&0.00946 (0.00590)\\
1.33364	&0.00434 (0.00039)	&0.00437 (0.00055)	&-0.00344 (0.00062)	&-0.00381 (0.00109)\\
1.33519	&0.00417 (0.00053)	&0.00478 (0.00084)	&-0.00288 (0.00083)	&-0.00004 (0.00174)\\
1.47077	&0.00273 (0.00105)	&0.00411 (0.00151)	&-0.00292 (0.00149)	&-0.00302 (0.00283)\\
1.48771	&-0.00011 (0.00179)	&-0.00035 (0.00295)	&-0.00368 (0.00239)	&0.00510 (0.00520)\\
1.57297	&0.00232 (0.00077)	&0.00360 (0.00105)	&-0.00110 (0.00091)	&0.00059 (0.00197)\\
1.57829	&0.00319 (0.00090)	&0.00326 (0.00155)	&-0.00203 (0.00125)	&-0.00076 (0.00287)\\
1.61351	&0.00231 (0.00048)	&0.00331 (0.00070)	&-0.00276 (0.00064)	&-0.00264 (0.00120)\\
\hline
\end{tabular}
\caption{Results for the strange electromagnetic form factors from the summation method and plateau fit at $1\,\text{fm}$ on ensemble N302.}
\end{table}

\end{document}